

\font\sectionfont=cmbx10 scaled\magstep1
\def\titlea#1{\vskip0pt plus.3\vsize\penalty-75
    \vskip0pt plus -.3\vsize\bigskip\bigskip
    \noindent{\sectionfont #1}\nobreak\smallskip\noindent}
\def\titleb#1{\medskip\noindent{\it#1}\qquad}
\def\claim#1#2{\vskip.1in\medbreak\noindent{\bf #1.} {\sl #2}\par
    \ifdim\lastskip<\medskipamount\removelastskip\penalty55\medskip\fi}
\def\rmclaim#1#2{\vskip.1in\medbreak\noindent{\bf #1.} {#2}\par
    \ifdim\lastskip<\medskipamount\removelastskip\penalty55\medskip\fi}
\def\beglemma#1#2\endlemma{\claim{#1 Lemma}{#2}}
\def\begdefinition#1#2\enddefinition{\claim{#1 Definition}{#2}}
\def\begtheorem#1#2\endtheorem{\claim{#1 Theorem}{#2}}
\def\begcorollary#1#2\endcorollary{\claim{#1 Corollary}{#2}}
\def\begremark#1#2\endremark{\rmclaim{#1 Remark}{#2}}
\def\begproposition#1#2\endproposition{\claim{#1 Proposition}{#2}}
\def\begassumption#1#2\endassumption{\claim{#1}{#2}}
\def\begProof{\noindent{\bf Proof.}\quad}
\def\begProofof#1{\medskip\noindent{\bf Proof of #1.}\quad}
\def\square{\hbox{$\sqcap\!\!\!\!\sqcup$}}
\def\endProof{\hfill\square\par
    \ifdim\lastskip<\medskipamount\removelastskip\penalty55\medskip\fi}
\newcount\FNOTcount \FNOTcount=1
\def\numfnot{\number\FNOTcount}
\def\addfnot{\global\advance\FNOTcount by 1}
\def\fonote#1{\footnote{$^\numfnot$}{#1}\addfnot}

\def\Co{{\bf C}} 
\def\Re{{\bf R}} 
\def\Ze{{\bf Z}} 

\def\A{{\cal A}}
\def\ua{{\rm C}^*({\cal A})}
\def\a{\alpha}
\def\B{{\cal B}}
\def\b{\beta}

\def\D{\Delta}

\def\f{\varphi}

\def\bG{{\bf G}}

\def\H{{\cal H}}

\def\I{{\cal I}}
\def\k{\kappa}

\def\l{\lambda}
\def\L{\Lambda}

\def\p{\pi}

\def\bP{{\bf P}}
\def\Q{\Omega}
\def\r{\rho}

\def\s{\sigma}

\def\th{\vartheta}

\def\w{\omega}
\def\x{\xi}

\def\z{\zeta}

\def\ad{{\rm ad}}

orthochronous
\def\imply{\Rightarrow}
\def\np{\par\noindent}
\def\npindent{\par}

\def\ov{\overline}


\def\refno#1#2{\item{[#1]}{#2}}
\def\begref#1#2{\titlea{#1}}
\def\endref{}
\def\np{\par\noindent}

\newcount\REFcount \REFcount=1
\def\numref{\number\REFcount}
\def\addref{\global\advance\REFcount by 1}
\def\wdef#1#2{\expandafter\xdef\csname#1\endcsname{#2}}
\def\wdch#1#2#3{\ifundef{#1#2}\wdef{#1#2}{#3}
    \else\write16{!!doubly defined#1,#2}\fi}
\def\wval#1{\csname#1\endcsname}
\def\ifundef#1{\expandafter\ifx\csname#1\endcsname\relax}

\def\autonumref{
    \def\rfr(##1){\wdef{q##1}{yes}\ifundef{r##1}$\diamondsuit$##1
        \write16{!!ref ##1 was never defined!!}\else\wval{r##1}\fi}
    \def\REF(##1)##2\endREF{\wdch{r}{##1}{\numref}\addref}\REFERENCES
    \def\references{
        \def\REF(####1)####2\endREF{
            \ifundef{q####1}\write16{!!ref. [####1] was never quoted!!}\fi
            \refno{\rfr(####1)}####2}
        \begref{References}{99}\REFERENCES\endref}}

\def\createbibfile{
    \def\REF(##1)##2\endREF{\wdch{r}{##1}{defined}}\REFERENCES
    \def\rfr(##1){\wdef{q##1}{quoted}\ifundef{r##1}$\diamondsuit$##1
    \write15{\string \REF(##1) [##1] to be inserted \csname endREF\endcsname}
        \write16{!!ref ##1 was never defined!!}\else##1\fi}
    \def\references{
        \def\REF(####1)####2\endREF{
            \ifundef{q####1}
            \else\refno{####1}{####2}
    \write15{\string \REF(####1)####2\csname endREF\endcsname}\fi}
    \begref{References}{99}\REFERENCES\endref}}

\def\REFERENCES{
 \REF(BiWi1) Bisognano J., Wichmann E., ``{\it On the duality condition for a
Hermitian scalar field}'', J. Math. Phys. {\bf 16} (1975),
985-1007.
 \endREF
 \REF(Borc1) Borchers H.J., ``{\it The CPT theorem in
two-dimensional theories of local observables}'', Commun. Math.
Phys. {\bf 143} (1992), 315.
 \endREF
 \REF(BrRo1) Bratteli O., Robinson D.W., `` Operator Algebras
and Quantum Statistical Mechanics, I.'' Sprin\-ger-Verlag,
Berlin-Heidelberg-New York, 1979.
 \endREF
 \REF(BGLo1) Brunetti R., Guido D., Longo R., ``{\it Modular
structure and duality in conformal quantum field theory}'', Commun.
Math. Phys., {\bf 156}  (1993), 201-219.
 \endREF
 \REF(BuFr1) Buchholz D., Fredenhagen K.,``{\it Locality and the
structure of particle states}'' Commun. Math. Phys. {\bf 84}
(1982), 1-54
 \endREF
 \REF(BuMT1) Buchholz D., Mach G., Todorov I.T.,``{\it The current algebra on
the circle as a germ of local field theories}'' Nucl. Phys. B
(Proc. Suppl.) {\bf 5B}, (1998) 20.
 \endREF
 \REF(DHRo1) Doplicher S., Haag R., Roberts J.E., ``{\it Local observables and
particle statistics I}'', Commun. Math. Phys. {\bf 23} (1971),
199-230.
 \endREF
 \REF(DHRo2) Doplicher S., Haag R., Roberts J.E., ``{\it Local observables and
particle statistics II}'', Commun. Math. Phys. {\bf 35} (1974),
49-85.
 \endREF
 \REF(Fred1) Fredenhagen K., ``{\it Generalization of the theory of
superselection sectors}'', in `The algebraic Theory of
Superselection Sectors', D. Kastler ed., World Scientific, Singapore
1990.
 \endREF
 \REF(FrJo1) Fredenhagen K., J\"or\ss\ M. ``{\it Conformal
Haag-Kastler nets, pointlike localized fields and the existence of
operator product expansion}'', DESY preprint, october 1994.
 \endREF
 \REF(FrRS1) Fredenhagen K., Rehren K.-H., Schroer B., ``{\it Superselection
sectors with braid group statistics and exchange algebras. I}'' Commum. Math.
Phys. {\bf 125} (1989) 201-226
 \endREF
 \REF(FrRS2) Fredenhagen K., Rehren K.-H., Schroer B., ``{\it Superselection
sectors with braid group statistics and exchange algebras. II}'' Rev. Math.
Phys. {\bf Special Issue} (1992) 113-157
 \endREF
 \REF(FrGM1) Fr\"ohlich J., Gabbiani F., Marchetti P., ``{\it Braid
statistics in three-dimensional local quantum theory}'' in `The algebraic
Theory of
Superselection Sectors', D. Kastler ed., World Scientific, Singapore
1990.
 \endREF
 \REF(GaFr1)  Fr\"ohlich J., Gabbiani F., ``{\it Operator algebras
and Conformal Field Theory}'',  Commun. Math. Phys. {\bf 155}
(1993), 569-640.
 \endREF
 \REF(GuLo1) Guido D., Longo R., ``{\it Relativistic invariance and charge
conjugation in quantum field theory}'', Commun. Math. Phys. {\bf
148} (1992), 521-551.
 \endREF
 \REF(GuLo2) Guido D., Longo R., ``{\it An algebraic spin and
statistics theorem}'', to appear in Commun. Math. Phys.
 \endREF
 \REF(Haag1) Haag R. , ``Local Quantum Physics'', Springer-
Verlag, Berlin-Heidelberg-New York 1992.
 \endREF
 \REF(HiLo1) Hislop P., Longo R., ``{\it Modular structure of the local
algebras
associated with the free massless scalar field theory}'', Commun. Math. Phys.
{\bf 84} (1982), 84.
 \endREF
 \REF(Jone1) Jones V., ``{\it Index for subfactors}'' Invent. Math.
{\bf 2} (1983) 1, 25.
 \endREF
 \REF(Long2) Longo R., ``{\it Index of subfactors and statistics of quantum
fields. I}'', Commun. Math. Phys. {\bf 126} (1989), 217-247.
 \endREF
 \REF(Long3) Longo R., ``{\it Index of subfactors and statistics of quantum
fields. II Correspondences, braid group statistics and Jones
polynomial }'', Commun. Math. Phys. {\bf 130} (1990), 285-309.
 \endREF
 \REF(Long4) Longo R., ``{\it Minimal index and braided subfactors}'' J. Funct.
Anal. {\bf 109} (1992), 98-112.
 \endREF
 \REF(Long5) Longo R., ``{\it Von~Neumann algebras and quantum
field theory}'' Proceedings of the International Congress of
Mathematicians, Zurich 1994, Birkh\"auser (to appear).
 \endREF
\REF(LoRo1) Longo R., Roberts J.E., ``{\it A theory of dimension}'',
preprint.
\endREF
 \REF(PiPo1) Pimsner M., Popa S. {\it Entropy and index for
subfactors} Ann. Sci. Ecole Norm. Sup. {\bf 19} (1986), 57-106.
 \endREF
 \REF(PrSe1) Pressley A., Segal G., `` Loop Groups'', Oxford
Science publ., Oxford 1986
 \endREF
 \REF(Robe1) Roberts J.E., ``{\it Local cohomology and superselection
structure}'', Commun. Math. Phys {\bf 51}, (1976)
107-119.
 \endREF
 \REF(StWi1) Streater R.F., Wightman A.S., ``PCT, Spin and
Statistics, and all that'', Addison Wesley, Reading (MA), 1989.
 \endREF
 \REF(Take1) Takesaki M., `` Tomita theory of modular Hilbert
algebras '', Lecture Notes in Math. {\bf 128}, Springer Verlag, New
York--Heidelberg--Berlin 1970.
 \endREF
 \REF(Take2) Takesaki M., ``Theory of Operator Algebras, I'',
Springer, New York, 1979.
 \endREF
 \REF(WWWi1) Wick G.C., Wightman A.S., Wigner E.P. ``{\it The
intrinsic parity of elementary particles}'' Phys. Rev. {\bf 88}
(1952), 101-105. \endREF
 \REF(Zimm1) Zimmer R. ``Ergodic Theory of Semisimple Groups'',
Birkh\"auser, Boston-Basel-Stuttgart, 1984.
 \endREF
}
\autonumref
%

\topskip3.cm
\font\ftitle=cmbx12 scaled\magstep1
\vskip2truecm
\centerline{\ftitle The Conformal Spin and Statistics Theorem}

 \bigskip\bigskip

\centerline{Daniele Guido$^1$\footnote{$^*$}
{ Supported in part by MURST and
CNR-GNAFA.} and Roberto Longo$^{1,2*}$}
\footnote{}{E-mail:\ guido@mat.utovrm.it,
longo@mat.utovrm.it }
 \vskip1.truecm \item{$(^1)$}
 Dipartimento di Matematica, Universit\`a di
Roma ``Tor Vergata'' \par via
della Ricerca Scientifica,
I--00133 Roma, Italia.
\item{$(^2)$} Centro Linceo Interdisciplinare,
Accademia Nazionale dei Lincei
\par via della Lungara 10, I--00165
Roma, Italia

\vfill\eject
\topskip0.cm

\titlea {Introduction} During the recent  years Conformal Quantum
Field Theory has become a widely studied topic, especially on a low
dimensional space-time,    because of physical motivations such as
the desire of a better understanding of two-dimensional critical
phenomena, and also for its rich mathematical structure providing
remarkable connections with different areas such as Hopf algebras,
low dimensional topology, knot  invariants, subfactors among many
others.

 The Operator Algebra approach furnishes a powerful tool of
investigation in this context, not only  because it naturally leads
to a model independent and intrinsic analysis, focusing on
essential aspects such as the relative position of the local von
Neumann algebras, but also because it makes visible  otherwise
hidden natural structures   bringing to   results  inaccessible by
different methods.

Two  examples of this kind,  the geometric description of the
Tomita-Takesaki modular structure of the local von~Neumann algebras
[\rfr(BiWi1),\rfr(HiLo1),\rfr(BGLo1)], and the connection of the
statistics of a superselection sector with the Jones index theory of
subfactors [\rfr(Long2)], will play a fundamental role in this
paper. These methods are present and important in general Quantum
Field Theory, but provide an even richer structure in the
low-dimensional case, conformal theories on $S^1$ in particular.

In the early seventies Doplicher, Haag and Roberts
[\rfr(DHRo1),\rfr(DHRo2)] developed a theory of superselection
sectors, in the sense of  [\rfr(WWWi1)], in the algebraic framework
proposed by Haag and Kastler [\rfr(Haag1)] starting from first
principles. They described a superselection sector by a localized
endomorphism $\r$ of the $C^*$-algebra generated by the local
observable von Neumann algebras on the usual Minkowski space. In
particular they showed that the  statistics of  $\r$, a
representation of the permutation group, is intrinsically encoded
in  $\r$ and classified it by an associated statistical parameter
$\l_\r$.

It  was  more recently realized that in the low dimensional case the
statistics becomes a representation of Artin braid group. By
applying  generalized DHR methods, a first analysis in this case
was given in  [\rfr(Long2),\rfr(FrRS1)]. In the
simplest cases (small index or few channels)
the statistics parameter classifies the braid group statistics  by
the Jones polynomial invariant for knots and links and its
generalizations, see [\rfr(Long3),\rfr(Long5)].

A key point in the analysis of superselection sectors is the
index-statistics theorem [\rfr(Long2)] showing that, in any
space-time dimension,
 $$
 {\rm Ind}(\r)=d(\r)^2
 $$
 where ${\rm Ind}(\r)$ is the minimal index of $\r$, an extension
of the Jones index [\rfr(Jone1)], and $d(\r):=|\l_\r|^{-1}$ is the
DHR statistical dimension of $\r$. We refer to [\rfr(Long5)] for a
survey and for references on the index theory for infinite factors,
but we recall that the square root of the minimal index of an
endomorphism of a factor has the meaning of a dimension, that find
an identification in this context by the above equation.

On the other hand important informations on the statistics are also
contained in the statistics phase $\k_\r:=\l_\r/|\l_\r|$ of
$\r$: on the  4-dimensional space-time  $\k_\r = \pm 1$, a sign
labeling the fundamental Fermi-Bose alternative. Therefore it is
natural to look at a counterpart of the index-statistics relation
for the statistics phase.

 Based on the classical spin-statistics connection (see
[\rfr(StWi1)]), one may easily conjecture that in a conformal theory
on $S^1$ the statistics phase has to agree with  the univalence  of
the sector $\r$
 $$
 s_\r=\k_\r
 $$
 where $s_\r:=e^{2\pi i L_\r}$ ($L_\r$ the conformal spin, the lowest
eigenvalue of the conformal Hamiltonian) is a label for the central
extension associated with the occurring projective representation of
the M\"obius group $PSL(2,\Re)$.

Attempts to prove this relation have been made in particular by
Fredenhagen, Rehren and Schroer [\rfr(FrRS2)] and,  in the related
$2+1$-dimensional context, by Fr\"ohlich, Gabbiani and Marchetti
[\rfr(FrGM1)]. Starting with assumptions on the existence of a global
conjugate charge and of complete reducibility, they obtained a  spin
summation rule, which implies the equality up to a sign
$s_\r=\pm\k_\r$. But the conformal spin-statistics theorem
remained unproven unless adding ad hoc undesirable assumptions.
Based on different ideas,  this paper will show how the full strength
of Operator Algebras  provides the general and intrinsic spin and
statistics relation, namely the  equality $s_\r=\k_\r$.
 We deal with conformal theories on $S^1$ (one-dimensional
components of two-dimensional chiral conformal theories) and base
our analysis only on first principles: isotony of the local von
Neumann algebras, locality, conformal invariance with positive
energy, existence of the vacuum. We thus obtain the complete
relation
 $$
{\rm statistics\  parameter}=
{{\rm univalence}\over{\sqrt{\rm minimal\ index}}}\ .
 $$

Note that $\k_\r$ has a local nature while $s_\r$ is a global
invariant. This  is reminiscent of familiar situations in Geometry
 and suggests that extensions of our result to more general (curved)
space-times should reveal further geometrical aspects. Our theorem
is not only a prototype for further generalizations, but it
already provides a number of immediate extensions or variants, like
for the case of  topological charges on a $2+1$-dimensional
space-time [\rfr(BuFr1)]. This is due to the fact that we shall  use
the conformal invariance only indirectly, not in an essential way.
For convenience we shall discuss these aspects together with related
points and examples in a separate paper.

 Our paper follows  a previous work [\rfr(GuLo1)] where we
reconsidered the classical spin and statistics theorem in Quantum
Field Theory [\rfr(StWi1)]  and derived it in the algebraic setting
assuming the  ``modular covariance property", namely the geometric
meaning of  the modular groups of the von~Neumann algebras
associated with wedge regions,  consistently to the
Bisognano-Wichmann theorem. That work, not directly extendible to
the lower dimensional case due to the occurrence of the braid group
statistics,  focused however on the role played by the modular
covariance property. The latter  was shown to hold   in conformal
field theory on  general grounds [\rfr(BGLo1),\rfr(GaFr1)], and
set thus the basis for the present analysis. Ultimalety only the
geometric description of the modular conjugations is essential in our
analysis.

 We now pass to the description of the more specific content of this
paper. In Section 1 we recall the basic properties shared by the
local von Neumann algebras $\A(I)$ associated with intervals $I$ of
$S^1$.

 Like in the classical case, the spin-statistics relation is
strictly tied up with the PCT symmetry. Section 2 is indeed devoted
to  the construction of a global conjugate charge for a
superselection sector $\r$ with finite statistics, a key point
relevant in itself, previously  an assumption  in the related
literature. As shown in [\rfr(GuLo1)], the sector
 $$\bar\r:=j\cdot\r\cdot j$$
 is locally a conjugate of $\r$ in the sense that if $\r$ is an
endomorphism localized in an interval $I_0$ and $j$ is the adjoint
(geometric) action given by the modular conjugation of an interval,
one has the identity
 $$ \bar\r|_{\A(I)}=\ov {\r|_{\A(I)}} $$
 where $I$ is any interval containing $I_0$ and its reflection by
$j$; the bar on the right hand side denotes the conjugate
endomorphism in the sense of the sectors of the factor $\A(I)$
[\rfr(Long3)], a framework equivalent to the setting of the
correspondences of Connes. In the irreducible case $\bar\r$ is
characterized by by the existence of  an isometry $V_I\in\A(I)$ that
intertwines the identity and $\bar\r\r|_{\A(I)}$. But the problem
remained whether there is a global intertwiner $V$ independent of
$I$. We solve this problem  positively by using an argument
inspired by the ``vanishing of the matrix coefficient theorem'' for
connected simple Lie groups, see the Appendix B.

We prove in fact the equivalence between the local and the global
intertwiners for superselection sectors with finite index, namely
the embedding into the sectors (endomorphisms modulo inners) of the
factor $M:=\A(I)$ determined,  via the restriction map,
 $$
 {\rm Superselection\, sectors}\rightarrow{\rm Sect}(M)
 $$
  corresponds by the index-statistics theorem to a faithful
functor of tensor $C^*$-categories with conjugates which is full (no
new intertwiner arises in the range). This implies that the fusion
rules of the superselection sectors are entirely described by the
 theory of subfactors.

As a first consequence we shall see in Section 3 that the (internal)
intertwiner property of the above isometry $V$ is equivalent to the
(spatial) property of being the standard implementation of $\r$,
according to  Araki, Connes and Haagerup, see Appendix A, with
respect to the vacuum vector. To extract information from this fact
we localize $\r$ in the upper-right quarter-circle and consider the
standard implementations $V_1$ and $V_2$ of $\r$ as an endomorphism
of the upper and  of the right semicircle von Neumann
algebra respectively and observe that
$$
\mu_\r:=V_1^*V_2^*V_1V_2 \eqno(3.1)
$$
is a scalar invariant for $\r$ that reflects by both
analytic-algebraic and geometric aspects.
 It is indeed  natural to look at $\mu_\r$ as a generalized
multiplicative commutator of local intertwiners, in the spirit of
the statistics, and identify it with the statistics parameter
$\l_\r$, or as an invariant obtained by reversing the orientation,
in the spirit of the spin, and identify $\mu_\r$ with ${\rm
Ind}(\r)^{-{1\over 2}}$--times the univalence of $\r$.

In more detail we shall obtain  the spin-statistics relation by
``squaring'' a more primitive identity between operators (see eq.
$(3.8)$) where a further invariant $c_\r$ enters. Our result is then
completed by showing that $c_\r$ is a conjugate-invariant character
on the semi-ring of the superselection sectors, so that it  takes
only the values $\pm 1$. This reaches the goal of our paper, but
leaves out the full understanding of the invariant $c_\r$, in
particular whether the value $c_\r=-1$ might actually occur. We
think this is the case, that reflects a cohomological obstruction,
and plan to return to this point somewhere else.

Our work has been announced in [\rfr(Long5)].

\titlea{1. General properties of conformal precosheaves on S$^1$}

In this section we recall the basic properties enjoyed by the family
of the von~Neumann algebras associated with a conformal Quantum Field
Theory on $S^1$.

By an  {\it interval} we shall always mean an open connected subset
$I$ of $S^1$  such that $I$ and the interior $I'$ of its complement
are non-empty. We shall denote by $\I$ the set of intervals in $S^1$.

A precosheaf $\A$ of von~Neumann algebras on the intervals of $S^1$
is  a map
 $$
 I\to\A(I)
 $$
 from $\I$ to the von~Neumann algebras on a Hilbert space $\H$ that
verifies the following property:

\begassumption{A. Isotony} If $I_1$, $I_2$ are  intervals and
 $I_1\subset I_2$, then
 $$
\A(I_1)\subset\A(I_2)\ .
 $$
\endassumption

$\A$ is a  {\it conformal precosheaf}  of von Neumann algebras if
the following properties B--E hold too.

\begassumption{B. Conformal invariance} There is a unitary
representation $U$ of $\bG$ (the universal covering group of
$PSL(2,\Re)$) on $\H$ such that
 $$
U(g)\A(I)U(g)^*=\A(gI)\ ,
\qquad g\in\bG,\ I\in\I.
 $$
 \endassumption

The group $PSL(2,\Re)$ is identified with the M\"obius group of
$S^1$, i.e. the group of conformal transformations on the complex
plane that preserve the orientation and leave the unit circle
globally invariant. Therefore  $\bG$ has a natural action on $S^1$.

\begassumption{C. Positivity of the energy} The generator of the
rotation subgroup $U(R(\cdot))$ is positive.
 \endassumption

Here $R(\th)$
denotes the (lifting to $\bG$ of the) rotation by an angle $\th$. In
the following we shall often write $U(\th)$ instead of $U(R(\th))$.
 We may associate two one-parameter groups with any
interval $I$.
 Let $I_1$ be the upper semi-circle, i.e. the interval
$\{e^{i\th},\th\in (0,\pi)\}$. We identify this interval with the
positive real line $\Re_+$ via the Cayley transform
$C:S^1\to\Re\cup\{\infty\}$ given by
 $z\to -i(z-1)(z+1)^{-1}.$
Then we consider the one-parameter groups $\L_{I_1}(s)$ and
$T_{I_1}(t)$ of diffeomorphisms of $S^1$ (cf. Appendix~B) such that
 $$
C\L_{I_1}(s)C^{-1}x=e^{s}x,\qquad
C T_{I_1}(t)C^{-1}x=x+t, \quad t,s,x\in\Re.
 $$
 We  also associate with $I_1$  the reflection $r_{I_1}$ given by
 $$
r_{I_1}z=\bar z
 $$
 where $\bar z$ is the complex conjugate of $z$.
 We remark that $\L_{I_1}$ restricts to an orientation preserving
diffeomorphisms of $I_1$, $r_{I_1}$ restricts to an orientation
reversing  diffeomorphism of $I_1$ onto $I'_1$ and $T_{I_1}(t)$ is
an orientation preserving  diffeomorphism of $I_1$ into itself if
$t\geq0$.

Then, if $I$ is an interval and we chose $g\in\bG$ such that
$I=gI_1$ we may set (see also the Appendix B)
 $$
\L_I=g \L_{I_1} g^{-1},\qquad
 r_I=g  r_{I_1} g^{-1},\qquad
 T_I=g  T_{I_1} g^{-1}.
 $$
 The elements $\L_I(s)$, $s\in\Re$ and $r_I$ are well defined,
while  the one parameter group $T_I$ is defined up to a scaling of
the parameter. However, such a scaling plays no role in this paper.
We note also that $T_{I'}(t)$ is an orientation preserving
diffeomorphism  of $I$ into itself if $t\leq0$.

Lemma B.4 in Appendix~B states the equivalence between the
positivity of the conformal Hamiltonian, i.e. the generator of the
rotation group $U(R(\cdot))$, and the positivity of the usual
Hamiltonian, i.e. the generator of the translations on the real line
in the above specified identification of $S^1$ with
$\Re\cup\{\infty\}$.

\begassumption{D. Locality} If $I_0$, $I$ are disjoint
intervals then  $\A(I_0)$ and $A(I)$ commute.
\endassumption

The lattice symbol $\vee$ will denote `the von Neumann algebra
generated by'.

\begassumption{E. Existence of the vacuum} There exists a unit
vector $\Q$ (vacuum vector) which is $U(\bG)$-invariant
 and cyclic for
$\vee_{I\in\I}\A(I)$. \endassumption

Let $r$ be an orientation reversing isometry of $S^1$ with $r^2=1$
(e.g. $r_{I_1}$). The action of $r$ on $PSL(2,\Re)$ by conjugation
lifts to an action $\s_r$ on $\bG$, therefore we may consider the
semidirect product of $\bG\times_{\s_r}\Ze_2$. Any involutive
orientation reversing isometry has the form $R(\th)r_{I_1}R(-\th)$,
thus $\bG\times_{\s_r}\Ze_2$ does not depend on the particular choice
of the isometry $r$. Since $\bG\times_{\s_r}\Ze_2$ is a covering of
the group generated by $PSL(2,\Re)$ and $r$, $\bG\times_{\s_r}\Ze_2$
acts on $S^1$. We call (anti-)unitary a representation $U$ of
$\bG\times_{\s_r}\Ze_2$ by operators on $\H$ such that $U(g)$ is
unitary, resp. antiunitary, when $g$ is orientation preserving,
resp. orientation reversing.

\begproposition{1.1} Let $\A$ be a conformal precosheaf. The
following  hold:
 \item{$(a)$}  {\it Reeh-Schlieder theorem} [\rfr(FrJo1)]: $\Q$
is cyclic and separating for each von Neumann algebra $\A(I)$,
$I\in\I$.
\item{$(b)$} {\it Bisognano-Wichmann property}
[\rfr(BGLo1),\rfr(GaFr1)]:  $U$ extends to an (anti-)unitary
representation of $\bG\times_{\s_r}\Ze_2$ such that, for any
$I\in\I$,
 $$
\eqalignno{U(\L_I(2\pi t))&=\D_I^{it}\ ,&(1.1)\cr
U(r_I)&=J_I\ ,&(1.2)\cr}
 $$
where $\D_I$, $J_I$ are the modular operator and the modular
conjugation associated with $(\A(I),\Q)$ [\rfr(Take1)].
 For each $g\in\bG\times_{\s_r}\Ze_2$
 $$
U(g)\A(I)U(g)^*=\A(gI)\ .
 $$
 \item{$(c)$} {\it Additivity} [\rfr(FrJo1)]: if a family of
intervals $I_i$ covers the interval $I$, then
 $$
\A(I)\subset\bigvee_i\A(I_i)\ .
 $$
  \item{$(d)$} {\it Spin and statistics for the vacuum sector}
[\rfr(GuLo2)]: $U$ is indeed a representation of $PSL(2,\Re )$,
i.e. $U(2\pi)=1$.
  \item{$(e)$} {\it Haag duality} [\rfr(BGLo1),\rfr(GaFr1)]:
$$\A(I')=\A(I)'\ ,\quad I\in\I.$$
 \endproposition
\begProof
We sketch here only the proof of $(d)$ and refer to the original
literature for the rest. Note however that:  the usual
Reeh-Schlieder argument shows that $(c)$ implies $(a)$;
$(b)$ is proved by using a theorem of Borchers [\rfr(Borc1)];
$(e)$ is an immediate consequence of $(b)$. To get $(d)$ let
$I_1$ and $I_2$ be  the upper and the right semicircle respectively,
then $J_{I_1}$ fixes $\Omega$ and implements an anti-automorphism of
$\A(I_{I_2})$, thus it commutes with $J_{I_2}$. By property $(b)$
$J_{I_1}J_{I_2}=U(\pi)$, thus $U(\pi)$ is an involution.
\endProof

 \begassumption{F. Uniqueness of the vacuum (or irreducibility)}
The only $U(\bG)$-invariant vectors are the scalar multiples of
$\Q$. \endassumption

The term irreducibility is due to the following.

 \begproposition{1.2} The following are equivalent:
 \item{$(i)$} $\Co\Q$ are the only $U(\bG)$-invariant vectors.
 \item{$(ii)$} The algebras $\A(I)$, $I\in\I$, are
factors. In this case they are type III$_1$ factors.
\item{$(iii)$} If a family of intervals $I_i$ intersects at only one
point $\z$, then $\cap_i\A(I_i)=\Co$.
\item{$(iv)$} The von~Neumann algebra $\vee\A(I)$ generated by
the local algebras coincides with $\B(\H)$ ($\A$ is
 irreducible).
\endproposition

 \begProof
$(i)\imply(ii)$. Indeed $(i)$ implies (c) of Corollary B.2 in the
Appendix B, hence the modular group of $\A(I)$ with respect to $\Q$
is ergodic, showing that $\A(I)$ is a type III$_1$ factor.
 \np
$(ii)\imply(iii)$. If $\z$ is a boundary point of an interval $I$,
then by additivity and duality $\cap_i\A(I_i)$ commutes both with
$\A(I)$ and $\A(I')$, and is therefore trivial.
 \np
$(iii)\imply(iv)$. We have
$\vee_{I\in\I}\A(I)\supset\vee_{\z\notin I}\A(I)=\B(\H)$.
 \np
$(iv)\imply(i)$. Let $I$ be an interval and $x\in\A(I)$ such
that $U(g)x\Omega=x\Omega$ for all $g\in\bG$. since $\Omega$ is
locally separating, we have $x=U(g)xU(g)^{-1}$. Since $\bG$ acts
transitively on the intervals,  $x$ is in the commutant of
$\cup_{I\in\I}\A(I)$, and is therefore a scalar. Since $\A(I)\Q$ is
dense in $\H$, by the Ergodic Theorem $\Q$ is the only
$U(\bG)$-invariant vector.
 \endProof

By Corollary~B.2 the irreducibility of $\A$ is also
equivalent to $\Q$ being unique invariant for any of the
one-parameter subgroups of $U$ corresponding to $T_I$,$\Lambda_I$ or
$R$.

Now any conformal precosheaf decomposes uniquely into a direct
integral of irreducible conformal precosheaves. This can be seen as
in Proposition 3.1 of [\rfr(GuLo2)]. We will therefore always
assume that our precosheaves are irreducible.

\titlea {2. Superselection structure. Constructing the global
conjugate charge}

\titleb {2.1 Generalities on superselection sectors with finite
index}

In this section $\A$ is an irreducible conformal precosheaf of
von Neumann algebras as defined in Section 1.

 A  covariant {\it representation} $\pi$ of  $\A$  is a family of
representations $\pi_I$ of the von Neumann algebras $\A(I)$,
$I\in\I$, on a Hilbert space $\H_\pi$ and a unitary representation
$U_\pi$ of the covering group $\bG$ of $PSL(2,\Re)$, with {\it
positive energy}, i.e. the generator of the rotation unitary
subgroup has positive generator, such that the following properties
hold:
 $$
\eqalign{
I\subset {\tilde I}\imply\pi_{\tilde I}\big|_{\A(I)}=\pi_I\quad
&(isotony)\cr
\ad U_\pi(g)\cdot\pi_I=\pi_{gI}\cdot\ad U(g) \quad
&(covariance).\cr} \eqno(2.1)
 $$
A unitary equivalence class of representations of $\A$
is called {\it superselection sector}.

Assuming $\H_\pi$ to be separable,  the representations $\pi_I$ are
 normal  because the $\A(I)$'s are factors [\rfr(Take2)]. Therefore
for any given $I_0$, $\pi_{I'_0}$ is  unitarily equivalent
$id_{\A(I'_0)}$ because $\A(I'_0)$ is a type III factor. By
identifying $\H_\pi$ and $\H$, we can thus assume that $\pi$ is
localized in a given interval $I_0\in\I$, i.e.
$\pi_{I'_0}=id_{\A(I'_0)}$ (cf. [\rfr(BuMT1)]). By Haag duality we
then have $\pi_I(\A(I))\subset\A(I)$ if $I\supset I_0$. In other
words, given $I_0\in\I$ we can choose in the same sector of $\p$ a
{\it localized endomorphism} with localization support in $I_0$,
namely a representation $\r$ equivalent to $\p$ such that
 $$
I\in\I,\,I\supset I_0\Rightarrow\r_I\in{\rm End}\A(I),
\qquad\r_{I_0'}=id_{I_0'}.
 $$
 In the following (with the exception of subsection 2.4)
representations or  endomorphisms are always assumed to be covariant
 with positive energy\fonote{Assuming strong
additivity (i.e. Haag duality on the real line) the  covariance
property with positive energy follows automatically in the finite
index case; in fact the weaker assumption of {\it 3-regularity} is
sufficient (cf. [\rfr(GuLo1)]). $\A$ is said to be $n$-regular if,
after removing $n$ points from $S^1$, the $C^*$-algebra generated by
the local operators is irreducible. By Haag duality and factoriality
any conformal precosheaf is  2-regular, but the validity of
3-regularity is not known in general. An example violating
4-regularity has been pointed out to us by H. Wiesbrock.}.

To capture the global point of view we may consider
the {\it universal algebra} $\ua$.
 Recall that
 $\ua$ is a C$^*$-algebra canonically associated with the
precosheaf $\A$ (see [\rfr(Fred1),\rfr(GuLo1)]). There are injective
embeddings $\iota_I:\A(I)\to\ua$ so that the local von Neumann
algebras $\A(I),\ I\in\I,$ are identified with subalgebras of $\ua$
and generate all together a dense $^*$-subalgebra of $\ua$, and
every representation of the precosheaf $\A$ factors through a
representation of $\ua$. Conversely any representation of $\ua$
restricts to a representation of $\A$. The vacuum representation
$\pi_0$ of $\ua$ corresponds to the identity representation of $\A$
on $\H$, thus $\pi_0$ acts identically on the local von Neumann
algebras. We shall often drop the symbols $\iota_I$ and $\p_0$ when
no confusion arises.

By the universality property, for each $g\in PSL(2,\Re)$ the isomorphism $\ad
U(g):\A(I)\to\A(gI)$, $I\in\I$ lifts to an automorphism $\a_g$ of $\ua$.
 It will be  convenient to lift the map $g\to\a_g$  to a representation, still
denoted by $\a$, of the universal covering group $\bG$ of $PSL(2,\Re)$ by
automorphisms of $\ua$.

The covariance property for an endomorphism $\r$ of $\ua$  localized
in $I_0$ means that $\a_g\cdot\r\cdot\a_{g^{-1}}$ is equivalent to
$\r$ for any $g\in\bG$, i.e.
 $$
\ad z_\r(g)^*\cdot\r=\a_g\cdot\r\cdot\a_{g^{-1}}\, \qquad
g\in\bG.\eqno(2.2)
 $$
for a suitable unitary $z_\r(g)\in\ua$. The map $g\to z_\r(g)$ can
be chosen to be a localized $\a$-cocycle, i.e.
 $$
\eqalign{
z_\r(g)&\in\A(I_0\cup gI_0)\quad \forall g\in\bG:I_0\cup
gI_0\in\I\cr z_\r(gh)&= z_\r(g)\a_g(z_\r(h)),\qquad
g,h\in\bG.\cr} \eqno(2.3)
 $$
 The relations between $(\pi,U_\pi)$ and $(\r,z_\r)$ are
 $$
\eqalign{
\pi&=\pi_0\cdot\r\cr
\pi_0(z_\r(g))&=U_\pi(g)U(g)^*\ .\cr}
\eqno(2.4)
 $$
 As is  known ([\rfr(Robe1)], see also [\rfr(GuLo1)]) that the
localized cocycle $z_\r$ reconstructs the endomorphism $\r$ via
$$
\r|_{\A(gI'_0)}=\ad z_\r(g)|_{\A(gI'_0)}\eqno(2.5)
$$
 A localized endomorphism of $\ua$ is said {\it irreducible} if the
associated representation $\pi$ is irreducible.

Note that  the representations $\pi_0\cdot\r_1$ and
$\pi_0\cdot\r_2$ associated with the endomorphisms $\r_1$, $\r_2$
of
$\ua$ are unitarily equivalent if and only if  $\r_1$ and $\r_2$ are
equivalent endomorphisms  of $\A$, i.e. $\r_2$ is a
perturbation of $\r_1$ by an inner automorphism of $\A$.

 An endomorphism of $\ua$ localized in an interval $I_0$ is said to
have {\it finite index} if $\r_I$ $(=\r|_{\A(I)})$ has finite index,
$I_0\subset I$ (see [\rfr(Long2),\rfr(Long5)]). The index is indeed
well defined due to the following.

\begproposition{2.1} Let $\r$ be an endomorphism localized in
the interval $I_0$. Then the  index ${\rm Ind}(\r):= {\rm
Ind}(\r_I)$, the minimal
index of $\r_I$, does not depend on the interval $I\supset I_0$.
\endproposition

 \begProof
We show indeed that all the inclusions $\r(\A(I))\subset \A(I)$
are isomorphic if $I\supset I_0$ (they are
isomorphic to the inclusion $\pi(\A(I))\subset\pi(\A(I'))'$ for all
$I\in\I$ ).
This follows because, if $g\in\bG$ and $z_\r(g)$ are chosen as in
$(2.2),(2.3)$ with $I\supset I_0$ and $gI = I_0$, then
 $$
\eqalign{
\{\r(\A(I_0))\subset\A(I_0)\}
&=     \{U_\r(g)\r(\A(I))U_\r(g)^*\subset U(g)(\A(I))U(g)^*\}\cr
&\simeq\{\r(\A(I))\subset z_\r(g^{-1})\A(I)z_\r(g^{-1})^*\}\cr}
 $$
 and $z_\r(g)\in \A(I)$.
\endProof
 \begproposition{2.2} Let $\r$ be a covariant (not necessarily
irreducible) endomorphism with finite index. Then the representation
$U_\r$ described before is unique. In particular, any irreducible
component of $\r$ is a covariant endomorphism.
 \endproposition
 \begProof
If $\r$ is localized in $I_0$ and has finite index the following
inclusion shows that $\pi(\ua)'$ is finite-dimensional,
$\pi:=\pi_0\cdot\r$:
 $$
\pi(\ua)'\subset (\pi(\A(I))\cup\pi(\A(I')))'=\pi(\A(I))'\cap\A(I)\ ,
\qquad I_0\subset I.\eqno(2.6)
 $$
 Since $U_\pi$ implements automorphisms of $\pi(\A)$, it implements
an  action of $\bG$ by automorphisms of $\pi(\A)'$, that must be
trivial because $\bG$ has no non-trivial action by automorphisms of
a finite-dimensional $C^*$-algebra. Indeed such an action should be
trivial on the center because $\bG$ is connected, thus it admits a
faithful invariant trace that defines a scalar product  unitarizing
the representation, but the only finite-dimensional unitary
representation of $\bG$ is the identity. Therefore we proved that
$U_\pi\in\pi(\ua)''$, and this fact implies that any irreducible
subsector of $\r$ is covariant.
 \np
Let $U'_{\pi}$ be another representation of $\bG$ as in (2.1).
 Then, for each $x\in\A(I)$, $I\in\I$,
 $$
U'_{\pi}(g)U_{\pi}(g)^*{\pi}(x)=U'_{\pi}{\pi}(\a_g(x))U_{\pi}(g)^*
={\pi}(x)U'_{\pi}(g)U_{\pi}(g)^*,
 $$
 which implies
$U'_{\pi}(g)U_{\pi}(g)^*$ to belong to
the center of $\pi(\ua)''$. Therefore
 $$
U_{\pi}(g)U'_{\pi}(g)^*U'_{\pi}(h)U_{\pi}(h)^*=
U'_{\pi}(g)U'_{\pi}(h)U_{\pi}(h)^*U_{\pi}(g)^*=
U'_{\pi}(gh)U_{\pi}(gh)^*\ ,
 $$
 i.e. $g\to U'_{\pi}(g)U_{\pi}(g)^*$ is a representation of $\bG$. Since $\bG$
is perfect, any abelian representation is trivial, i.e. $U_{\pi}=U'_{\pi}$
 \endProof

By the above proposition  the {\it univalence} of an
endomorphism $\r$ is well defined by
 $$
s_\r = U_{\r}(2\pi)\ .
 $$
 By definition $s_\r$ belongs to $\pi(\ua)'$ therefore, when $\r$ is
irreducible, $s_\r$ is a complex number of modulus one
$$s_\r=e^{2\pi iL_\r}$$
with $L_\r$ the lowest weight of $U_\r$. In this
case, since  $U_{\r'}(g):=\pi_0(u)U_\r(g)\pi_0(u)^*$, where
$\r'(\cdot):=u\r(\cdot)u^*$, $u\in\ua$, then $s_\r$ depends
only on the superselection class of $\r$.

Let $\r_1$, $\r_2$ be  endomorphisms of an algebra
$\cal B$.  Their intertwiner space is defined by
 $$
(\r_1,\r_2)= \{T\in{\cal B}: \r_2(x)T=T\r_1(x),\ x\in{\cal
B}\}.\eqno(2.7)
 $$
 In case ${\cal B} =\ua$, $\r_i$ localized in the interval
$I_i$  and $T\in (\r_1,\r_2)$, then $\pi_0(T)$ is an intertwiner
between the representations $\pi_0\cdot\r_i$. If $I\supset I_1\cup
I_2$, then by Haag duality its embedding  $\iota_I\cdot \pi_0(T)$ is
still an intertwiner in $(\r_1,\r_2)$ and a local operator.
We shall denote by $(\r_1,\r_2)_I$ the space of such local
intertwiners
$$
(\r_1,\r_2)_I= (\r_1,\r_2)\cap\A(I).
$$
If $I_1$ and $I_2$ are disjoint, we may cover $I_1\cup I_2$ by an
interval $I$ in two ways: we adopt the convention that, unless
otherwise specified, a {\it local intertwiner} is an element of
$(\r_1,\r_2)_I$ where $I_2$ follows $I_1$ inside $I$ in the
clockwise sense.

We now define  the statistics. Given the endomorphism $\r$ of $\A$
localized in $I\in\I$, choose an equivalent endomorphism $\r_0$
localized in an interval $I_0\in\I$ with $\bar I_0\cap\bar I
=\emptyset$ and let $u$ be a local intertwiner in $(\r,\r_0)$ as
above, namely $u\in (\r,\r_0)_{\tilde I}$ with $I_0$ following
clockwise $I$ inside $\tilde I$.

The {\it statistics operator} $\varepsilon := u^*\r(u) =
u^*\r_{\tilde I}(u) $ belongs to $(\r^2_{\tilde I},\r^2_{\tilde
I})$.  An elementary computation shows that it gives rise to a
presentation of  the Artin  braid group
$$\varepsilon_i\varepsilon_{i+1}\varepsilon_i =
\varepsilon_{i+1}\varepsilon_i\varepsilon_{i+1},\qquad
\varepsilon_i\varepsilon_{i'} =\varepsilon_{i'}\varepsilon_i
\,\quad{\rm if}\,\, |i-i'|\geq 2,$$ where
$\varepsilon_i=\r^{i-1}(\varepsilon)$. The (unitary equivalence
class of the) representation of the braid group thus obtained is the
{\it statistics} of the superselection sector $\r$.

Recall that if $\r$ is an endomorphism of a C$^*$-algebra $\B$,
a {\it left inverse} of $\r$
 is a completely positive map $\Phi$ from $\B$ to itself  such that
$\Phi\cdot\r=id$.

We shall see in Corollary 2.12 that  if $\r$ is irreducible there
exists a unique left inverse $\Phi$ of $\r$ and that the
{\it statistics parameter}
$$
\l_\r:=\Phi(\varepsilon)\eqno(2.8)
$$
  depends only
on the sector of $\r$.

 The
{\it statistical dimension} $d(\r)$ and the  {\it statistics phase}
$\k_\r$ are then defined by
 $$
d(\r) = |\lambda_\r|^{-1}\ ,\qquad \k_\r = {\lambda_\r
\over |\lambda_\r|}\ .
 $$
 We shall indeed prove  the equality between the statistics phase
and the univalence while the statistical dimension equals the square
root of the index [\rfr(Long2)] (see Corollary 3.7).

\titleb {2.2  Equivalence
between local and global intertwiners}

If $\r,\s$ are  endomorphisms of  $\ua$ localized in the
interval $I$,  we may consider their intertwiner space
$(\r_I,\s_I):=\{T\in \A(I):\s(x)T=T\r(x), \, \forall x\in \ua\}$.
We always have $(\r,\s)_I\subset(\r_I,\s_I)$.

\begtheorem{2.3} Let $\r$, $\s$ be endomorphisms with
finite index localized in $I_0$. Then
 $$
(\r_I,\s_I)=(\r,\s)_I
 $$
for any $I\in\I$ that contains $I_0$. In other words if
$T\in(\r_I,\s_I)$ then $\iota_I(T)$ intertwines $\r$ and $\s$ in
$\ua$ \endtheorem

The proof of this theorem will be carried on in a few steps. In the
following $\r$ denotes an endomorphism of $\ua$ with finite
index localized in an interval $I_0$.   Let  $\z\in I_0'$
and identify $S^1\backslash\z$ with $\Re$. Then $\r$ restricts  to an
endomorphism  of  each von Neumann algebra $\A(-\infty,\ell)$, for
sufficiently large $\ell\in\Re$, hence it gives rise to an
endomorphism $\r_\z$  the $C^*$-algebra
$\A_\z$, the norm closure of $\cup_{\ell\in\Re}\A(-\infty,\ell)$.
Let $\bP$ be the stabilizer of the point $\z$ for the $PSL(2,\Re)$
action, namely the semidirect product of the translations $T(t)$ and
dilations $\L(s)$ on $\Re$:  each $g\in\bP$ is written uniquely
as a product $g=T(t)\L(s)$. Notice that $\bP$ is canonically embedded
in $\bf G$ since $\bP$ is simply connected and its Lie algebra is a
subalgebra of the Lie algebra of $PSL(2,\Re)$ that coincides with
the Lie algebra of $\bf G$. It follows that $U_\r$ restricts to a
representation of $\bP$ and we set
 $$
\b_g(x)=U_\r( g)xU_\r(
g)^*=z_\r(g)U(g)xU(g)^*z_\r(g)^*,\quad x\in \A_\z,\,g\in \bP,
 $$
so that $\b$ is an action of $\bP$ by automorphisms of $\A_\z$, due
to the fact that the cocycle $z_\r$ is a local operator.

We consider now the semigroup $\bP_0$, the semidirect product of
negative dilations with  positive translations. $\bP_0$ is an
amenable semigroup and we need an invariant mean $m$ constructed as
follows: first we average (with an invariant mean) on positive
translations and then over negative dilations.
 Observe that  $f\to\int_{\bP_0}f(g)dm(g)$  gives an invariant mean
on all $\bP$ vanishing on  $f$ if, for any given $s\in\Re$, the map
$t\to f(T(t)\L(s))$  has support in a left half line.

 Then we associate to $m$ the completely positive map $\Phi_\z$
of $\A_\z$ to $\B(\H)$ given by
 $$
 \Phi_\z(x):=\int_{\bP_0} z_\r(g)^*
xz_\r(g)dm(g), \quad x\in \A_\z. \eqno(2.9)
 $$

\beglemma{2.4}  $\Phi_\z$ is a  left inverse
of $\r_\z$. Moreover $\Phi_\z$ is locally normal, i.e. has  normal
restriction to $\A(-\infty,\ell)$, $\ell\in\Re$, and
$\bP$-invariant, namely
 $$
\Phi_\z=\a_g^{-1}\Phi_\z\b_g,\quad g\in\bP.
 $$
 \endlemma

 \begProof Let $x$ belong to $\A(-\infty,\ell)$, $\ell\in\Re$. By
formula $(2.2)$
 $$
\Phi_\z(\r_\z(x))=\int_{\bP_0} \a_g(\r_\z(\a_{g^{-1}}(x)))dm(g)
=x
 $$
because of the above property of $m$ since the integrand is
constantly equal to $x$ on the set $\{g\in\bP_0$ :
$g^{-1}(-\infty,a)$ $\cap$ $I_0$ $=$ $\emptyset\}$.
 Then the localization of $\r_\z$ and Haag duality imply that
the range of $\Phi_\z$ is contained in $\A_\z$.

Setting $E=\r_\z\cdot\Phi_\z$ we have a conditional expectation
of $\A_\z$ onto the range of $\r_\z$ that restricts to a
conditional expectation $E_{(-\infty,\ell)}$ of $\A(-\infty,\ell)$ onto
$\r(\A(-\infty,\ell))$ if $(-\infty,\ell)\supset I_0$. Since
$\r_{(-\infty,\ell)}$ is assumed to have finite index,
$E_{(-\infty,\ell)}$ is automatically normal [\rfr(Long3)].
Therefore $\Phi_\z|_{\A(-\infty,\ell)}=
\r_{(-\infty,\ell)}^{-1}E_{(-\infty,\ell)}$ is
normal for $\ell$ large enough, hence for any $\ell$.
 \np
Concerning the $\bP$-invariance of $\Phi_\z$ we have, making use of
the cocycle condition,
 $$
\eqalign{
\a_g^{-1}\Phi_\z\b_g(x)
&=\a_g^{-1}( \int_{\bP_0} z_{\r}(h)^*\b_g(x)z_\r(h)dm(h))\cr
&=\a_g^{-1}(\int_{\bP_0}z_\r(h)^*z_\r(g)
  \a_g(x)z_\r(g)^*z_\r(h)dm(h))\cr
&=\int_{\bP_0} z_\r(hg^{-1})^*x z_\r(hg^{-1})dm(h)=\Phi_\z(x)\cr}
 $$
\endProof

\begcorollary{2.5}  $\f = \omega\Phi_\z$ is a locally
normal $\b$-invariant state  on $\A_\z$, where $\omega =
(\ \cdot\ \Omega, ,\Omega)$. \endcorollary \begProof
 We have
$\f\b_g=\omega\Phi_\z\b_g=\omega\a_g\Phi_\z=\omega\Phi_\z
=\f$ and $\f$ is locally normal because both
$\omega$ and $\Phi_\z$ are locally normal.
\endProof

Let $\{\p_\f,\x_\f,\H_\f\}$ be the GNS triple
associated with the above state $\f$ and $V$ be the unitary
representation of $\bP$ on $\H_\f$ given by $V_g
x\x_\f=\b_g(x)\x_\f$ for $x\in\A_\z$. Notice that
$V$ is strongly continuous because $\f$ is locally normal. We now
need a variation of known results, see [\rfr(DHRo2),\rfr(BuFr1)].

 \beglemma{2.6} If $\r_\z$ is irreducible
then
 $$
\f(x)
=\int_{\bP_0}\b_g(x)dm(g),\quad
x\in\A_\z
 $$
\endlemma

\begProof
If $x\in\A(-\infty,\ell)$ and $y\in\A_\z$ is localized in a bounded
interval, the commutator function
$t\to[\b_{T(t)\L(s)}(x),\r(y)]
=\b_{T(t)\L(s)}([x,\r(\a_{T(t)\L(s)}^{-1}(y)])$
vanishes on a right half line, hence
$[\int_{\bP_0}\b_g(x),\r(y)dm(g)]=\int_{\bP_0}[\b_g(x),\r(y)]dm(g)=0$.

 Since $\r_\z$ is locally normal, $\int_{\bP_0}\b_{g}(x)dm(g)$
commutes with every $\r(\A(-\infty,\ell))$, thus with
$\r_\z(\A_\z)$;
 since $\r_\z$ is irreducible it therefore a scalar equal to its
vacuum expectation value:
 $$
\int_{\bP_0}\b_g(x)dm(g)= \int_{\bP_0}\omega(\b_g(x))dm(g)
=\int_{\bP_0}\omega(z_g^*xz_g)dm(g)=\omega\Phi_\z(x)=\f(x),
$$
due to the fact that $\omega$ is normal and $\a$-invariant. \endProof

\begcorollary{2.7} If $\r_\z$ is irreducible, the one-parameter
(translation) unitary group $V(T(t))$  has
positive generator.
 \endcorollary

 \begProof
If $f\in L^1(\Re)$   has Fourier transform $\hat
f$ with support in $(-\infty,0)$, we have to show that
$V_f:=\int_\Re f(t)V(T(t))dt=0$. Choose by Lemma B.4 a non zero
$\psi\in\H$ such that ${\rm sp}_{U_\r}(\psi)+{\rm supp}\hat f\subset
(-\infty,0)$, where ${\rm sp}_{U_\r}$ denotes the spectrum relative
to $U_\r(T(\cdot))$. Setting $\b_f:=\int_{\Re}f(t)\b_{T(t)}dt$, for
any $x\in\A_\z$ the vector $\b_g(\b_f(x))\psi=0$, for all $g\in
\bP_0$, since it has negative spectrum relative to $U_\r(T(\cdot))$.
By averaging over $\bP_0$ the vector $\b_g(\b_f(x^*)\b_f(x))\psi$,
Lemma 2.6 implies $\|V_fx\x_\f\|^2=\f(\b_f(x)^*\b_f(x))=0$.
 \endProof

 \begcorollary{2.8} If $\r_\z$ is irreducible, $\f$ is faithful
on $\cup\r(\A(-\infty,\ell))$.
 \endcorollary

 \begProof
$\A_\z$ is a simple $C^*$-algebra since it is the inductive limit
of type III factors (that are simple $C^*$-algebras). Therefore
$\p_\f$ is one-to-one and the statement will follow if we show
that $\x_\f$ if cyclic for
$\B_\ell:=\r(\A(-\infty,\ell))',\,\ell>0$. To this end we may use a
classical Reeh-Schlieder argument. If $\psi\in\H$ is orthogonal to
$\B_\ell\x_\f$, and $\ell_0>\ell$, then for all
$x\in\B_{\ell_0}$ we have $(x\x_\f,V(T(t))\psi)=0$ for $t$ in a
neighborhood of $0$, thus for all $t\in\Re$ by positivity of the
generator shown by Corollary 2.7.
 Hence, setting $\a_t\equiv\a_{T(t)}$ and $\b_t\equiv\b_{T(t)}$,
$\psi$ is orthogonal to $(\cup_t \b_t(\B_{\ell_0}))\x_\f$, thus
$\psi=0$ because $\cup_t \b_t(\B_{\ell_0})$ is irreducible since
 $$
\eqalign{
(\bigcup_t\b_t(\B_{\ell_0}))'
&=\bigcap_t \b_t(\r(\A(-\infty,\ell_0))=
\bigcap_t\r(\a_t(\A(-\infty,\ell_0)))\cr
 & = \r(\bigcap_t
\a_t(\A(-\infty,\ell_0))) =\bigcap_\ell \A(-\infty,\ell)= \Co\cr} $$
 by the local normality of $\r$. \endProof

 \begproposition{2.9} $(\r_I,\r_I)$ does not
depend on the interval  $I\supset I_0$.
\endproposition
\begProof
 We begin with the case in which $\r_\z$ is irreducible and assume
for convenience that $\bar I_0\subset (-\infty,0)$. Notice then that
 $(\r_{(-\infty,0)},\r_{(-\infty,0)})$ is finite-dimensional
and, by covariance, globally $\b_g$-invariant with $g$ in the
subgroup of dilations because these transformations preserve
$(-\infty,0)$.  Therefore
$(\r_{(-\infty,0)},\r_{(-\infty,0)})\x_\f$ is a
finite-dimensional subspace of $\H_\f$ globally invariant for
$V(\L(s))$, $s\in\Re$. By Proposition B.3 of the appendix B we thus
have $V(T(t))x\x_\f=x\x_\f$ for every element $x\in
(\r_{(-\infty,0)},\r_{(-\infty,0)})$, thus
$\b_{T(t)}(x)=x$ because $\x_\f$ is separating.
 It follows that if $x\in(\r_{(-\infty,0)},\r_{(-\infty,0)})$ and
$y\in \A(-\infty,0)$
 $$
[x,\r(\a_g(y))]=\b_g([\b_g^{-1}(x),\r(y)])=\b_g([x,\r(y)])=0
 $$
namely
 $$
x\in (\r_{(-\infty,0)},\r_{(-\infty,0)})\Rightarrow x\in
(\r_\z,\r_\z)=\Co I
 $$
Since the converse implication is obvious by Haag duality we have
the equality of the two intertwiner spaces.

	Now if $\r$ is any
endomorphism with finite index,  $(\r_\z,\r_\z)$ is
finite-dimensional by the inclusion $(2.6)$,  and $\r_\z$ decomposes
into a direct sum of irreducible endomorphisms of $\A_\z$ which are
covariant by Proposition~2.2, therefore the preceding analysis shows
that also in
this case $(\r_{(-\infty,0)},\r_{(-\infty,0)})=(\r_\z,\r_\z)$.
 Since $(\r_\z,\r_\z)$ is translation invariant, we get
$(\r_{(-\infty,\ell)},\r_{(-\infty,\ell)})=(\r_\z,\r_\z)$
whenever $I_0\subset(-\infty,\ell)$ and, since $\z$ was arbitrary,
we get the thesis.
 \endProof

\begProofof{Theorem 2.3} The case $\s=\r$ follows immediately by
Proposition 2.6: if $T\in (\r_I,\r_I)$ then $T$ also belongs to
$(\r_{\tilde I},\r_{\tilde I})$ for any interval $\tilde I\supset
I$ hence by additivity $T$ is a self-intertwiner of $\r$ on
the whole algebra $\ua$.

To handle the general case consider  a  direct sum endomorphism
$\eta:=\r\oplus\s$ localized in $I$, then
$$
{\rm dim}(\eta_I,\eta_I)={\rm dim}(\r_I,\r_I)+{\rm
dim}(\s_I,\s_I) +2{\rm dim}(\r_I,\s_I)
$$
while
$$
{\rm dim}(\eta,\eta)_I={\rm dim}(\r,\r)_I+{\rm
dim}(\s,\s)_I +2{\rm dim}(\r,\s)_I
$$
therefore ${\rm dim}(\r_I,\s_I)={\rm dim}(\r,\s)_I$ and since we
always have $(\r,\s)_I\subset (\r_I,\s_I)$ these two intertwiner
spaces coincide.
 \endProof

In particular we have proved the following.

\begcorollary{2.10} Let $\r$ be an endomorphism of $\ua$ with finite
index localized in $I_0$. The following are equivalent:
\item{$(i)$} $\p_0\cdot\r$ is an irreducible representation of $\ua$
\item{$(ii)$} $\r(\A(I))'\cap\A(I)=\Co$ for some, hence for all,
$I\supset I_0$
\item{$(iii)$} $\r_\z(\A_\z)'\cap \A_\z =\Co$
\item{$(iv)$} $\r_\z$ is an irreducible representation of $\A_\z$.
\npindent Moreover any finite index representation $\pi$ of $\ua$ is
the direct sum of irreducible representations.
 \endcorollary

 \titleb
{2.3 The  conjugate sector}

Let $\r$ be an endomorphism of $\ua$ with finite index and
localized in the interval $I_0$ as before. We shall say that
the endomorphism $\bar\r$ is a conjugate of $\r$ if there
exist isometries $V\in (id,\bar\r\r)$ and $\bar V\in (id,\r\bar\r)$
such that
$$
\bar V^*\bar\r(V)={1 \over d},\quad V^*\r(\bar V)={1 \over
d}\eqno(2.10)
 $$
where $d$ is a positive scalar. In this case one can in fact choose
$V,\bar V$ so that $d$ is the square root of the minimal index of
$\r$.

Denote by $j_I$ the lifting to an anti-automorphism of $\ua$ of the
adjoint action of the modular conjugation $J_I$ on the precosheaf
$\A$.
\begtheorem{2.11} Let $\r$ be a covariant endomorphism with
finite index. There exists a conjugate endomorphism $\bar\r$,
unique as superselection sector.  $\bar\r$ is covariant with
positive energy and is given by the formula
$$
\bar\r=j\cdot\r\cdot j \eqno(2.11)
$$
where $j=j_I$.
If both $\r$ and $\bar\r$ are localized in the interval $I$,
then there exist isometries $V\in (id,\bar\r\r)_I$ and
$\bar V\in (id,\r\bar\r)_I$ such that the conjugate
equations $(2.10)$ holds with $d=\sqrt{{\rm Ind}(\r)}$.
 \npindent
 If moreover $\r$ is irreducible, then $\bar\r$ is the unique
irreducible endomorphism of $\ua$, up to inner automorphisms, such
that $\r\bar\r$ contains the identity  and in this
case there exists a unique (up to a phase) isometry
$V\in(id,\bar\r\r)_I$.
 \endtheorem

 \begProof
As shown in [\rfr(GuLo1)], $\bar\r :=j_I\cdot\r\cdot j_I$ is an
endomorphism of $\ua$ locally conjugate to $\r$, namely
 $\bar\r_{\tilde I}$ is a conjugate endomorphism of
$\r_{\tilde I}$ according to [\rfr(Long3)], for any interval $\tilde
I$ such that both $\r$ and $\bar\r$ are localized in $\tilde I$.
Fixing such an interval $\tilde I$, since $\r_I$ has finite
index,  there exist isometries $V\in(id_{\tilde I},\bar \r_{\tilde
I}\r_{\tilde I})$, $\bar V\in(id_{\tilde I},\r_{\tilde I}\bar
\r_{\tilde I})$ such that $\bar V^*\bar\r(V)={1 \over d}$,
$V^*\r(\bar V)={1 \over d}$, with $d=\sqrt{{\rm Ind}(\r_{\tilde
I})}$ [\rfr(Long3)]. By Theorem 2.3 $V$ and $\bar V$ are global
intertwiners, namely $\bar \r$ is a global conjugate. The uniqueness
of $\bar\r$,  the characterization of $\bar\r$ in the irreducible
case and the uniqueness of $V$ follow again by the corresponding
statements for  sectors of factors [\rfr(Long3)] because of Theorem
2.3. The covariance of $\bar \r$ follows by the formula
$\bar\r=j\cdot\r\cdot j$, see [\rfr(GuLo1)].
 \endProof

 \begcorollary{2.12} If $\r$ is a an endomorphism of $\ua$ with
finite index, there exists a (global) faithful left inverse $\Phi$
of $\r$ which is given by the formula
$$
\Phi = V^*\bar\r(\cdot)V \eqno(2.12)
 $$
where $V\in(id,\bar\r\r)$ verifies the conjugate equations
$(2.10)$ and all faithful left inverses have this form.
If $\r$ is localized in $I$, also $\Phi$ is localized in $I$ and
$\Phi|_{\A({\tilde I})}$ is normal if $\tilde I\supset I$.
\npindent
 If $V$ $\bar V$ are chosen so that the constant $d$ in $(2.10)$ is
equal to $\sqrt{{\rm Ind}(\r_I)}$, then $\Phi$ is uniquely
determined. In particular if $\r$ is irreducible  then $\Phi$ is the
unique left inverse of $\r$. \endcorollary
 \begProof
Only the uniqueness of $\Phi$ needs
still to be proved. We  assume that $\r$ is localized in $I$ and
$V\in\A(I)$. By the same argument as in Corollary 5.7 of
[\rfr(Long2)], essentially the push-down lemma in [\rfr(PiPo1)],
every element $x\in\ua$ can be written as
 $$
x={\rm Ind}(\r)\r\Phi(x\bar V^*)\bar V.\eqno(2.13)
$$
If $\Psi$ is a left inverse of $\r$ and satisfies the conjugate
equations with $d=\sqrt{{\rm Ind}(\r)}$, then $\Psi$ and $\Phi$ have
the same restriction to $\A(I)$ because the corresponding
statement is true for endomorphisms of factors [\rfr(Long3)]
and, by Corollary~2.10, $\Psi(\bar V)=\Phi(\bar V)$. Thus, by
formula $(2.13)$,
 $$
\Psi(x)={\rm Ind}(\r)\Phi(x\bar V^*)\Psi(\bar V)=
{\rm Ind}(\r)\Phi(x\bar V^*)\Phi(\bar
V)=\Phi(x).
 $$
\endProof

If $\r$ is a finite index endomorphism of $\ua$, we define
$\l_\r=\Phi(\varepsilon)$ where $\Phi$ is the unique ``minimal" left
inverse provided by Corollary 2.12. As shown in [\rfr(Long2)],
$\Phi$ is a standard left inverse in the sense of [\rfr(DHRo2)],
namely $\l_\r$ is a positive scalar multiple of a unitary
$\k_\r\in(\r,\r)_I$ and the statistical dimension is then defined by
$d(\r)=\|\l_\r\|^{-1}$. By the index-statistics theorem (see
Corollary 3.7) if $\r$ has finite index, then also $d(\r)$ is finite.

\begcorollary{2.13} If $\r$ is irreducible with finite index, the
statistics parameter $\l_\r$ in formula $(2.8)$ is a non-zero scalar.
\endcorollary

\begProof $\l_\r=\Phi(\varepsilon)=\Phi_\z(\varepsilon)$ belongs to
$(\r_I,\r_I)$ thus is a scalar by Corollary 2.10. $\l_\r$ does not
vanish as mentioned above. \endProof

\titleb{2.4 Equivalence between finite index and finite statistics.}

If a covariant, positive energy superselection sector $\r$ has
finite index, then also the statistical dimension is finite. In
fact  Corollary 3.7 will relate the two quantities in the general
reducible case.  For completeness, in this subsection we will
outline an argument showing a converse of this assertion. We shall
say  that  a  localized endomorphism $\r$ of $\ua$ has {\it finite
statistics} if there exists a left inverse $\Phi$ of $\r$ such that
the statistical parameter $\l_\r:=\Phi(\varepsilon)$ is an
invertible operator; even in the irreducible case we do not know a
priori that $\l_\r$ is a scalar since
 Corollary 2.10 has not been proved.

In the following proposition $\r$ is a covariant endomorphism of
$\ua$, but positive energy is not assumed.

\begproposition{2.14} If $\r$ is covariant with finite statistics,
then $\r$ has  finite index and positive energy.  \endproposition

\begProof Let $\r$ be localized in $I_0$, $\z\in I'_0$ and
$\Phi_\z:=\Phi|_{\A_\z}$. Because of finite statistics  the DHR
inequality holds:
 $$
 \|\Phi_\z(x)\|\geq c\|x\|,\quad x\in\A_\z^+\eqno(2.14)
 $$
 where $c=\|\l_\r^{-1}\|^{-2}>0$, by reasoning as in [\rfr(DHRo2)].
Indeed if $x=x^*\in\A(-\infty,\ell)$ with $I_0$ $\subset$
$(-\infty,\ell)$ and $u$ is a unitary such that $u\r(\cdot)u^*$
is localized in $(\ell,\infty)$, so that $\r(x)=u^*xu$ and
$\varepsilon$ $=$ $u^*\r(u)$, we have $\Phi(u^*x)$ $=$ $\Phi(\r(x)u^*)$
$=$ $\Phi(\r(x)\varepsilon\r(u^*))$ $=$ $x\l_\r u^*$ and therefore
$\|\Phi(x^2)\|$ $\geq$ $\|\Phi(xu)\Phi(u^*x)\|$ $=$
 $\|\l_\r^*x^2\l_\r\|$ $=$ $\|x\l_\r^*\l_\r x\|$ $=$ $\geq c\|x^2\|$.

 As $\r_\z$ is isometric, the inequality (2.14) is clearly
equivalent to the Pimsner-Popa inequality [\rfr(PiPo1)]
 $$
 \|E(x)\|\geq c\|x\|,\quad x\in\A_\z^+,
 $$
 with $E=\r_\z\cdot\Phi_\z$ the associated conditional expectation
onto the range of $\r_\z$, and it is also equivalent to
 $$
 E(x)\geq cx,\quad x\in\A_\z^+,\eqno(2.15)
 $$
 (see [\rfr(Long2)] for the  version of these inequalities on
infinite factors). In particular $E|_{\A(I)}$ is normal and $\r_I$
has finite index $I\supset I_0$.

We can now replace $\Phi_\z$ by its average $\Phi'_\z$ over $\bP$
with respect to an invariant mean, e.g. the $m$ in the previous
section, $\Phi'_\z:=\int_{{\bP}_0}\a_g^{-1}\Phi_\z\b_gdm(g)$.
Since $\r$ is locally normal $\r\Phi'_\z$ still satisfies the
inequality $(2.15)$ and hence $\Phi'_\z$ the inequality $(2.14)$.

At this point  the state $\varphi=\omega\Phi'_\z$ in Corollary 2.5 is
again locally normal and faithful, thus Proposition 2.9 applies and
provides the global conjugate in Theorem 2.11. The usual additivity
of the spectrum argument then shows that $\r$ is a positive energy
representation. \endProof

\titlea {3. The conformal spin-statistics theorem}

\titleb {3.1 A first relation between spin and statistics}

In this subsection we prove a first relation between spin and
statistics. We shall not use the full conformal invariance, but only
the covariance with respect to the rotation subgroup and the
geometric interpretation of the modular conjugations.

In the following  $I_1$ and $I_2$  always denote the upper
semicircle  $\{e^{i\theta},\th\in (0,\pi)\}$ and the right
semicircle  $\{e^{i\th},\th\in (-{\pi\over 2},{\pi\over 2})\}$
respectively and $\r$ is an irreducible covariant with positive
energy endomorphism of $\ua$ with finite  index localized in an
interval whose closure is contained in $I_1\cap I_2$. Then
$\r_{I_i}=\r |_{\A(I_i)}$ is an irreducible finite-index
endomorphism of $\A(I_i)$  and we denote by $V_i$ the standard
implementation of $\r_{I_i}$, $i=1,2$, with respect to the vacuum
vector, see Appendix A. We also shorten the notations: $J_i$ stands
for the modular conjugation $J_{I_i}$, $\ov\r_i$ for the conjugate
$j_i\r j_i$ of $\r$ where $j_i$ is the promotion to an
anti-automorphism of $\ua$ of the precosheaf anti-automorphism
$J_i \cdot J_i$. The symbol $\ad U$  denotes the
automorphism of $\ua$  corresponding to a unitary $U$ (e.g.
$\ad U_\r(g):=\ad z_\r(g)\cdot\alpha_g$).

\beglemma{3.1}  We have $V_1\in\A(I_2)$, $V_2\in\A(I_1)$ and $V_i$
is the unique isometry (up to a phase) with this localization
support that intertwines  the identity  and $\r\ov\r_i$, $i=1,2$.
\endlemma

\begProof  By the geometric meaning of $J_1$, both $\r$ and $\ov\r_1$
are  localized in $I_2$, thus by Theorem 2.11  we
can take an isometry $v\in (id,\ov\r_1\r)_{I_2}$, in fact $v$
belongs to $\A(I)$ if $I$ is any subinterval of $I_2$ that contains
both the localization support of $\r$ and of $\ov\r_1$. Since
$\r_{I_2}$ is irreducible, $v$ is uniquely determined (up to a
phase) by such properties. Therefore we may choose $v$ so that
$j_1(v)=v$. By additivity  $v$ implements $\r_{I_1}$ and since it
also commutes with $J_1$ we have $V_1=\pm v$  by Lemma A.3. The
argument for $V_2$ is similar. \endProof

Since $\r,\bar\r_1,\bar\r_2$ are localized in disjoint intervals,
they pairwise commute, thus $V_1V_2$ and $V_2V_1$ both belong to
$(id,\r^2\ov\r_1\ov\r_2)_{I_1\cup I_2}$ hence
 $$
 \mu_\r=V_1^*V_2^*V_1V_2
 $$
 is a scalar. It is an invariant for $\r$ that, by construction,
reflects algebraic, analytical and geometric aspects. By looking  at
$\mu_\r$ from these different point of view we shall identify it,
with different arguments, with the statistics parameter and with
the univalence of $\r$ times $d(\r)^{-1}$, proving the conformal
spin-statistics theorem.

\beglemma{3.2}  The following identities between endomorphisms of
$\ua$ hold:
 \item{$(a)$} $\r\ov\r_1=\ad U_\r(\pi)\r\ov\r_2\ad U(\pi),$
 \item{$(b)$} $\r\ov\r_1j_2\r\ov\r_1j_2=\r\ov\r_2j_1\r\ov\r_2j_1.$
\endlemma
\begProof By formula $(1.2)$ we have  $J_1J_2=U(\pi)$, hence
 $j_1j_2=j_2j_1=\ad U(\pi)$, therefore
$$
\ov\r_1=\ad U(\pi)\ov\r_2\ad U(\pi).
 $$
Thus by covariance
 $$
\r\ov\r_1=\r\ad U(\pi)\ov\r_2\ad U(\pi)=
\ad U_\r(\pi)\r\ov\r_2\ad U(\pi)
 $$
 and, since $\ov\r_1$ and $\ov\r_2$ are localized in
disjoint intervals and thus commute,
 $$
\r\ov\r_1j_2\r\ov\r_1j_2=\r\ov\r_1\ov\r_2\ad U(\pi)\r\ad U(\pi)
=\r\ov\r_2\ov\r_1\ad U(\pi)\r\ad U(\pi)=\r\ov\r_2j_1\r\ov\r_2j_1
 $$
\endProof

\beglemma{3.3} We have
 $$
 U_\r(\pi)V_2U(\pi)=c_\r V_1\eqno(3.2)
 $$
where $c_\r$ is a complex number of modulus one. \endlemma

\begProof By Lemma 3.1 $V_1$ is the unique isometry (up to a phase)
in $(id,\r\ov\r_1)_{I_2}$. By Lemma 3.2 $(a)$, also
$U_\r(\pi)V_2U(\pi)$ belongs to $(id,\r\ov\r_1)$. Moreover, if $x\in
\A(I'_2)$, then ad$U(\pi)(x)\in\A(I_2)$ hence
 $$
 {\ad}U_\r(\pi)V_2U(\pi)(x)={\ad}U_\r(\pi)\r{\ad}U(\pi)(x)
 ={\ad}U_\r(2\p)\r(x)=\r(x)=x
 $$
showing that $U_\r(\pi)V_2U(\pi)$ belongs to $\A(I_2)$ too,
thus it coincides with $V_1$ up to a phase.
 \endProof

 \beglemma{3.4}
$\b_\r:=(V_1J_2V_1J_2)^*V_2J_1V_2J_1
 $ belongs to $(0,1]$.
 \endlemma

 \begProof
According to Lemma~3.2 $ (ii)$,  $V_1J_2V_1J_2$
and $V_2J_1V_2J_1$   are both isometries in $(id,
\r\ov\r_1j_2\r\ov\r_1j_2)$ and  both belongs to the same  local
von Neumann algebra $\A(I)$ where $\r\ov\r_1j_2\r\ov\r_1j_2$ is
localized in $I\in\I$
 therefore  $\b_\r$ is a complex scalar.

Setting $e_i:=V_iV_i^*$ we deduce that
 $$
\b_\r V_1J_2V_1J_2=e_1J_2e_1J_2\ V_2J_1V_2J_1.\eqno(3.3)
 $$

Since $V_i$ is the standard implementation of $\r_{I_i}$, $V_i$
preserves
the positive cone ${\cal P^{\natural}}(\A(I_i),\Q)$.
Moreover $J_1$ preserves ${\cal P^{\natural}}(\A(I_2),\Q)$
because it
implements an antiautomorphism of $\A(I_2)$ and fixes $\Omega$, and
$J_2$ preserves ${\cal P^{\natural}}(\A(I_1),\Q)$ analogously.
 By the definition of
the natural positive cones and the relations $V_1,e_1\in\A(I_2)$,
$V_2\in\A(I_1)$, we have that $V_2J_1V_2J_1\Q$ and $V_1J_2V_1J_2\Q$
belong to ${\cal P^{\natural}}(\A(I_1),\Q)\cap
{\cal P^{\natural}}(\A(I_2),\Q)$ and
$e_1J_2e_1J_2\Q\in{\cal P^{\natural}}(\A(I_2),\Q)$.

Since the scalar product of non-zero vectors in a natural cone
is non-negative, and furthermore positive if one of the vectors is
cyclic (equivalently separating), and since
$(e_2J_1e_2J_1\Q,\Q)$ $=\|\D_{I_1}^{1\over 4}e_2\Q\|^2\neq 0$ we
have
 $$(V_1J_2V_1J_2\Q,\Q)>0,\quad
(e_2J_1e_2J_1V_2J_1V_2J_1\Q,\Q)=(V_2J_1V_2J_1\Q,e_2J_1e_2J_1\Q)>0$$
that entails $\b_\r>0$ by comparing with (3.3),  provided we show
that $V_2J_1V_2J_1\Q$ is separating for $\A(I_1)$. But this is true
because if $x\in\A(I_1)$ and $xV_2J_1V_2J_1\Q=0$ then
$$(j_1\Phi j_1\Phi(x^*x)\Q,\Q)=
(J_1V^*_2J_1V^*_2x^*xV_2J_1V_2J_1\Q,\Q)=0$$
and this implies $x=0$ because the left inverse $\Phi$ of $\r$ is
faithful.
  The rest is clear since by definition
$||\b_\r||\leq 1$.  \endProof

 \beglemma{3.5}$
\l_\r=V_1^*V_2^*V_1V_2\ .
 $
 \endlemma
 \begProof As in [\rfr(DHRo1)] we get $\l_\r=\r(V_1^*)V_1;$ indeed
if $\r'$ is localized in $I_1\cap I'_2$ and $u$ is a
unitary in $(\r,\r')_{I_1}$, then $\ad u^*|_{\A(I_2)}=\r_{I_2}$,
thus  $\r(V_1^*)V_1 = u^*V_1^*uV_1 = u^*\Phi(u) = \Phi
(\varepsilon_\r) = \l_\r$.
 Since
$V_1\in\A(I_2)$ and $V_2$ implements $\r$ on $\A(I_2)$ we thus have
 $$
V_1^*V_2^*V_1V_2=V_1^*\Phi(V_1)=\Phi(\r(V_1^*)V_1)=\Phi(\l_\r)=
\l_\r.\eqno(3.5)
 $$\endProof
 \begproposition{3.6} The
following relations hold:
 $$
\eqalignno{
\b_\r&=d(\r)^{-1} &(3.6)\cr
s_\r&=c_\r^2 \k_\r&(3.7)\cr}
 $$
 where $\k_\r$ is the phase of the statistical parameter.
 \endproposition
 \begProof
Taking adjoints in (3.2), we have
$U(\pi)V^*_1U_\r(\pi)=c_\r V_2^*$, and multiplying side by side
this expression with formula (3.2) we have
 $$c_\r^2
V_2^*V_1= s_\r U(\pi)V_1^*V_2U(\pi)\eqno(3.8)
 $$
because $s_\r:=U_\r(2\pi)$.

Since $J_i$ commutes with $V_i$ and $J_1J_2=J_2J_1=U(\pi)$ we
have
 $$
\b_\r=J_1\b_\r J_1=J_1(V_1^*J_2V_1^*J_2V_2J_1V_2J_1)J_1
=V_1^*U(\pi)V_1^*V_2U(\pi)V_2\eqno(3.9)
 $$
therefore, by inserting formula (3.8) in the expression for $\l_\r$
given by
Lemma~3.5 and comparing with (3.9) we obtain
 $$
\l_\r=V_1^*V_2^*V_1V_2=c_\r^{-2} s_\r
V_1^*U(\pi)V_1^*V_2U(\pi)V_2=c_\r^{-2}
s_\r\b_\r\eqno(3.10)
 $$
 and the thesis easily follows.
 \endProof
 \begcorollary{3.7} (Index-statistics theorem) For every
covariant endomorphism $\r$ of $\ua$ we have
${\rm Ind}(\r)=d(\r)^2.$
\endcorollary
\begProof  If $\r$ is irreducible we have $\bar V_1^*\r_{I_2}(V_1)$
$=$ ${1\over d}$ with $d=\sqrt{{\rm Ind}(\r)}$ by Corollary 2.12 and
comparing with formula $(3.5)$ we have the thesis since $\bar V_1$
and $ V_1$ are equal up to a phase. The general case follows  by
additivity of both the statistical dimension and the square root of
the minimal index  (or by a direct  argument). The case of infinite
index is treated in subsection 2.4.\endProof

\titleb{3.2 The spin-statistics theorem.}

We prove now that $c^2_\r =1$, completing our result. In this step
the role of the conformal invariance is to fix uniquely the
representation of the rotation group $U_\r(\th)$, otherwise defined
up to a one-dimensional representation, as the restriction of the
unique representation of $\bG$. We could nevertheless fix $U_\r(\th)$
by using the positivity of the conformal Hamiltonian.

It is convenient to extend the definition of $c_\r$  to the
case of a  reducible  finite index $\r$. To this end notice that, as
in the proof Lemma 3.3, both
$U_\r(\pi)V_2U(\pi)$ and $c_\r V_1$  belong to
$(id,\r\bar\r_1)_{I_2}$, thus there exists
$c_\r  \in(\r\bar\r_1,\r\bar\r_1)_{I_2}$ such
that formula (3.2) holds. Replacing $c_\r$ by its push-down if
necessary, we may further assume that $c_\r  \in(\r,\r)$ and this
condition define it uniquely, see [\rfr(LoRo1)].

In the following $\r$, $\s$ are finite index endomorphisms of $\ua$.
	\beglemma{3.8} Let $\r$ and  $\s$  be localized in
$I_1\cap I_2$, with $\r$ an irreducible subsector of $\s$ and
$p_\r\in\A(I_1\cap I_2)$ is the minimal idempotent in
$(\s,\s)_{I_1\cap I_2}$ corresponding to $\r$, then $c_\s p_\r=c_\r
p_\r$.  In particular, if $c_\s$
is a scalar, then $c_\r=c_\s.$
\endlemma
\begProof
With $w\in
(\r,\s)$ an isometry in $\A(I_1\cap I_2)$, we have by Lemma A.4 of
the appendix
$$
\sqrt{d(\r)}V_i^\r=\sqrt{d(\s)}w^*Jw^*JV_i^\s.\eqno(3.11)
$$
 The projection $p_\r=w^*w\in\A(I_1\cap I_2)$ commutes with the
range of $\s$, hence it commutes with $U_\s$ (see the proof
of Proposition 2.2), therefore
 $$(w^*U_\s(g)w)(w^*U_\s(h)w)=w^*U_\s(gh)w,\quad
g,h\in{\bf G},\eqno(3.12)$$ namely $g\to w^*U_\s(g)w$ is a unitary
representation. Since for every $x\in\ua$ we have
 $$\eqalign{
(w^*U_\s(g)w)\r(x)(w^*U_\s(g)^*w)
&=w^*U_\s(g)\s(x)U_\r(g)^*w\cr
&=w^*\s(U(g)xU(g)^*)w=\r(U(g)xU(g)^*),\cr}
 $$
 we get by the uniqueness of the representation in Proposition
2.2
 $$U_\r(g)=w^*U_\s(g)w.\eqno(3.13)$$
 Since $c_\s$ leaves in a finite-dimensional algebra, we may
assume that $p_\r$ is an eigen-projection of $c_\s$ namely
$c_\s p_\r =\ell p_\r$ with $\ell\in\Co$.
Making
substitutions in the formula $(3.2)$ according to the equations
$(3.12)$, $(3.13)$, we then get
 $$\eqalign{\sqrt{d(\r)}c_\r V^\r_1&= \sqrt{d(\s)}
w^*U_\s(\p)ww^*J_2w^*J_2V_2^\s U(\p)\cr
&=\sqrt{d(\s)}w^*U_\s(\p)J_2w^*J_2V_2^\s U(\p)\cr &=
\sqrt{d(\s)}w^*
U_\s(\p)J_2w^*J_2U_\s(-\p)c_\s V_1^\s \cr &=
\sqrt{d(\s)}
w^*U(\p)z_\s(\p)J_2w^*J_2z^*_\s(\pi)U(\p)c_\s V_1^\s\cr &=
\sqrt{d(\s)} w^*U(\p)J_2w^*J_2U(\p)c_\s V_1^\s\cr
&=\sqrt {d(\s)}
w^*J_1w^*J_1 c_\s V_1^\s\cr
&=\sqrt{d(\s)} w^* c_\s J_1w^*J_1V_1^\s =\sqrt{d(\r)} \ell V^\r_1
\cr} $$
where we have used that $[J_2w^*J_2,z_\s(\p)]=0$ due to the
localization in disjoint intervals of $w^*$ and $z_\s((\pi))$ and
again of the identity $J_1J_2=U(\pi)$, and
this concludes the proof.
\np\endProof
Our choice of the intervals $I_1$ and $I_2$ is, of course,
conventional. If we replace them by their rotates $R(\th)I_1$,
$R(\th)I_2$, we would get a priori another invariant
$c_\r(\th)$ for a $\r$ localized in their intersection.
But this
is soon seen to be equal to $c_{\r_\th}$, the old invariant for
$\r_\th := {\rm ad}U(-\th)\r{\rm
ad}U(\th)={\rm ad}z_\r(-\th)\r$ (because $U(\th)$ establishes
an isomorphism between the old and the rotated structures). Next
lemma implies that $c_{\r_\th}=c_\r$ if also $\r_\th$ is
localized in $I_1\cap I_2$.
\beglemma{3.9}
$c_\r$ depends only on the superselection class of  $ \r$ and
not on its representative $\r$ nor on the choice of $I_1$ and $I_2$
as above. \endlemma
\begProof
If $\r$ is localized in $I_1\cap I_2$ and $\s={\rm ad}W^*\cdot\r$
for some unitary $W\in\A(I_1\cap I_2)$ then $V_i^\r=W^*JW^*JV_i^\s$
and by a computation similar to the one in the  Lemma 3.7 we
see that $c_\s=c_\r$. By the comment preceding the this
lemma it thus follows that $c_\r$ remains unchanged if we rotate the
$I_i$'s provided $\r$ stays localized in the intersection of the
intervals. Thus, in finitely many steps, replacing $\r$ by an
equivalent endomorphism and making small rotations of the intervals,
we see that $c_\r$ does not vary in its superselection class.
\endProof
\beglemma{3.10}  $c_\r=c_{\bar\r}$.
\endlemma
\begProof
By Lemma 3.9 we may choose $\bar\r=\bar\r_1=j_1\r j_1$. Thus
$\bar\r$ is localized in $I_1'\cap I_2$ and $c_{\bar\r}$ is
definable with respects to the intervals $I_2=R(-{\pi\over
2})I_1$,$\,I_1'=R(-{\pi\over 2})I_2$. The standard implementations of
$\bar\r$ relative to these intervals are respectively given by
$J_1V_2^\r J_1$ and $J_1V_1^\r J_1=V_1^\r$, moreover
$U_{\bar\r}(\th)=J_1U_\r(-\th)J_1$, see [\rfr(GuLo1)]. Inserting these
identities in the defining expression $(3.2)$ for $c_{\bar\r}$ we
thus have
$J_1U_\r(-\pi)J_1V_1^\r U(\pi)=c_{\bar\r}J_1V_2^\r J_1$ and after
cancellations this gives  the stated equality.
\endProof \beglemma{3.11}
  Let $\r$, $\s$ be irreducible and localized in $I_1\cap I_2$.
 Then   $c_{\r\s}=c_\r  c_\s$.
\endlemma
\begProof
 By the cocycle equation $z_{\r\s}(g)
=z_{\r}(g)\r(z_{\s}(g))$ and the multiplicativity of the standard
implementations
$V_i^{\r\s}=V_i^{\r}V_i^{\s}$
 the equation
$(3.2)$ for $\r\s$ gives
$$\eqalign{
c_\r c_\s U(\p)V_1^\r V_1^\s U(\p)
&=z_\r(\p)V_2^\r z_\s(\p) V_2^\s
=z_\r(\p)\r(z_\s(\pi))V_2^\r V_2^\s\cr
&=z_{\r\s}(\p)V_2^{\r\s}
=c_{\r\s}U(\p)V_1^{\r\s}U(\p)\cr}\eqno(3.14)$$
where  we used that $z_\s(\p)\in\A(I_2\cup I_1')$
and that $V_2^\r$ implements $\r$ on $\A(I_2\cup I_1')$.
 Since $V_1^{\r\s}=V_1^\r V_1^\s$ we have the thesis.
\endProof
\begcorollary{3.12}  $c_\r ^2=1$.
\endcorollary
\begProof
 If $\r$ is irreducible, then by the Lemmas 3.8 and 3.9 we have
$c_\r^2=c_\r c_{\bar \r}=
 c_{\r{\bar \r}}= 1$. The general case follows by Lemma 3.8.
\endProof
  Now the spin and statistics relation immediately follows
immediately by Proposition 3.6.
 \begtheorem{3.13} (Spin and Statistics) Let $\r$ be a
superselection sector with finite statistics. Then $
\k_\r=s_\r.$
 \endtheorem
\titlea{Appendix A. Standard implementation of left inverses}

We will deal here with  the the notion of standard
implementation  (see e.g. [\rfr(BrRo1)]) in
the endomorphism case.

Let $M$ be a von~Neumann algebra on a Hilbert
space $\H$ and $\r$ a unital injective endomorphism of $M$.  The left
inverses $\Phi$ of $\r$ correspond bijectively to the conditional
expectations $E$ of $M$ onto $\r(M)$:
 $$
\eqalign{
\Phi&\to E=\r\cdot\Phi\cr
E&\to\Phi=\r^{-1}\cdot E\,.\cr}
\eqno(A.1)
 $$
We shall say that an isometry $V\in\B(\H)$ {\it implements the
left inverse}
  $\Phi$
if
 $$
V^*xV=\Phi(x)\,,\qquad x\in M\eqno(A.2)
 $$

\beglemma{A.1} Let the isometry $V$ implement $\Phi$. Then
\item{$(a)$} $Vx=\r(x)V$, \quad $x\in M,$
\item{$(b)$} $exe=E(x)e$,  \quad $x\in M,$
\np
where $e=VV^*$ and $E=\r\Phi$.
 Conversely if $(a)$ and $(b)$ hold then $V$ implements $\Phi$.
\endlemma

\begProof
 If we set $x\equiv\r(y)$ in $(A.2)$ we have
 $V^*\r(y)V=y$ for all $y\in M$
hence
 $$
e\r(y)V=Vy.\eqno(A.3)
 $$
In particular, if $y$ is unitary,
 $\|e\r(y)V\x\|$ $=$ $\|Vy\x\|$ $=$ $\|\x\|$ $=$ $\|\r(y)V\x\|$,
$\x\in\H$, showing that $e\r(y)V\x=\r(y)V\x$, hence
$e\r(y)e=\r(y)e$, so we have
 $$
e\r(y)=(\r(y^*)e)^*=(e\r(y^*)e)^*=e\r(y)e=\r(y)e
 $$
which implies $e\in\r(M)'$ because $M$ is generated by its
unitaries. Formula $(A.3)$ then entails $(a)$.
 To check $(b)$ notice that
 $$
exe=VV^*xVV^*=V\Phi(x)V^*=
=\r(\Phi(x))VV^*=E(x)e.
 $$
 Conversely, assuming $(a)$ and $(b)$, we have
 $$
V^*xV=V^*exeV=V^*E(x)V=V^*\r(\Phi(x))V=\Phi(x),\quad x\in M\ .
 $$
 \endProof
We shall say that an isometry $V$ {\it implements the
endomorphism $\r$} and that the projection $e$ {\it implements the
conditional expectation} $E$ if the equations $(a)$ and $(b)$ of
Lemma A.1 are respectively satisfied.

We now fix a unit cyclic and separating vector $\Q\in\H$ for $M$ and
its corresponding natural cone ${\cal P}^{\natural}(M,\Q)$.

If $\Phi$ is a normal left inverse of $\r$ let us consider the state
 $$
 \f=\w\cdot\Phi
 $$
where $\w=(\cdot\Q,\Q)$ and the corresponding vector $\x\in
{\cal P}^{\natural}(M,\Q)$ such that $\f=(\cdot\x,\x)$.

Let $e:=[\r(M)\x]\in\r(M)'$ and let $V_\Phi$ be the isometry of $\H$
with final projection $e$ such that $V_\Phi :\H\to e\H$ is the
Araki-Connes-Haagerup standard implementation of $\r$ as an
isomorphism of $M$ with $\r(M)$ with respect to the positive cones
${\cal P}^{\natural}(M,\Q)$ and  ${\cal P}^{\natural}(\r(M),\x)$.
Then $V_\Phi$ is given by
 $$
 V_{\Phi}x\Q=\r(x)\x\,,\qquad x\in M
 $$
We check that $V_\Phi$ implements  $\Phi$. To this end note first
that $E=\r\Phi$ is $\f$-invariant since
 $$
\f\cdot E=\w\cdot\Phi\cdot\r\cdot\Phi=\w\cdot\Phi=\f
 $$
 Then
 $$
\eqalign{
(x\r(b)\x,\r(a)\x)&=\f(\r(a^*)x\r(b))=\f\cdot E(\r(a^*)x\r(b))=\cr
&=\f(\r(a^*)E(x)\r(b))=(E(x)\r(b)\x,\r(a)\x)\quad a,b,x\in M\cr}
 $$
i.e. $eE(x)e=exe$, $x\in M$, but $e\in\r(M)'$, hence $e$ implements
$E$; in particular, if $\Phi$ is faithful, $e$ is the Takesaki
projection for $E$.

Moreover $V_\Phi$ implements $\Phi$ because
 $$
\r(x)V_\Phi y\Q =\r(x)\r(y)\x
=\r(xy)\x=V_\Phi xy\Q\,
\quad x,y\in M.
 $$
The isometry $V_\Phi$ will be called the {\it standard implementation}
of $\Phi$ with respect to  $\Q$. In case $\r$ has finite index,
namely $\r(M)$ is a finite index subfactor of $M$, and $\Phi =
\Phi_{\rm min}$, the minimal left inverse of $\r$, we shall denote
$V_{\Phi_{\rm min}}$ by $V_\r$ and call it the standard
implementation of $\r$ with respect to $\Q$.

 We collect here some properties of the standard implementations.

\begproposition{A.2}
 \item {$(a)$} $V_\Phi$ is the unique isometry that implements $\Phi$
and  sends ${\cal P}^{\natural}(M,\Q)$ into itself. In particular
$V_\Phi$ depends on ${\cal P}^{\natural}(M,\Q)$ but not on the
particular vector $\Q$.
 \item {$(b)$} $V_\Phi$ is the unique isometry that implements $\Phi$
and verifies $V_\Phi\Q\in {\cal P}^{\natural}(M,\Q)$.
 \item {$(c)$} $V_{\Phi_1\Phi_2}=V_{\Phi_1}V_{\Phi_2}$, with
$\Phi_1$, $\Phi_2$ normal left inverses of $\r_1$, $\r_2$. In
particular, if $\r_1$, $\r_2$ have finite index,
$V_{\r_1\r_2}$ $=$ $V_{\r_1}V_{\r_2}$.
 \item {$(d)$} $JV_\Phi J=V_\Phi$, where $J$ is the modular
conjugation of $(M,\Q)$.
 \endproposition
\begProof
 By construction $V_\Phi$  implements $\Phi$ and maps ${\cal
P}^{\natural}(M,\Q)$ into itself, in particular $V_\Phi\Q\in {\cal
P}^{\natural}(M,\Q)$.  Now suppose that an isometry $V$ implements
$\Phi$ and $V\Q\in {\cal P}^{\natural}(M,\Q)$.
 Then
 $$
(\ \cdot\ V\Q,V\Q)=(V^*\cdot V\Q,\Q)=\w\cdot\Phi=\f
 $$
thus $V\Q$ is the unique vector $\x\in{\cal
P}^{\natural}(M,\Q)$ associated with $\f$ and
$Vx\Q=\r(x)V\Q=\r(x)\x$, namely $V=V_\Phi$ . This proves
$(a)$ and $(b)$.
 \np
$(c)$ is consequences of $(a)$ and of the
multiplicativity of the minimal index [\rfr(Long4)].
 \np
 $(d)$ $J$ restricted to the range of $e=V_\Phi V_\Phi^*$ coincides
with the modular conjugation of $M_e$ because $\varphi$ preserves
the conditional expectation $E$, thus $V_\Phi^*JV_\Phi =J$ because
$V_\Phi$ is the standard implementation of $\r$ as an isomorphism of
$M$ with $\r(M)$.
 \endProof
\beglemma{A.3} Let $M$ be a factor and $\r$ a finite index
endomorphism. If $W$ is an isometry that implements $\r$ and
commutes with $J$, then $W$ implements a  left inverse
$\Phi$ of $\r$ and $W=mV_\r$ for some  $m\in (\r,\r)$, which is
invertible iff $\Phi$ is faithful. In particular, if
 $\r$ is irreducible, then $W=\pm V_\r$
\endlemma
\begProof The partial isometry $Z=WV^*_\r$ commutes with $J$
and belongs to  $\r(M)'$, thus $Z\in N'\cap M_1$, where we set
$N=\r(M)$ and $M_1=JN'J$ denotes the Jones basic extension of
$N\subset M$. Clearly we have $W = ZV_\r$. Let $m$ be the
Pimsner-Popa push-down of $Z$, namely the unique element $m\in M$
such that $mV_\r = ZV_\r$. We have $m={\rm Ind}(\r)E(Ze)$ with
$E=\r\Phi_{\rm min}$, thus $m\in \r(M)'\cap M$ and $W=mV_\r$ showing
in particular that $W$ implements a left inverse $\Phi$ of $\r$.
Clearly $\Phi$ is faithful if $m$ is invertible. Conversely, if
$\Phi$ is faithful, $p\in(\r,\r)$ is a projection and $pm=0$ then
$\Phi(p)=V_\r^*m^*pmV_\r =0$, thus $p=0$ so $m$ is invertible.

If moreover $\r$ is irreducible, then $m\in \Co$, thus $m=\pm 1$
because both $W$ and $V_\r$ are isometries commuting with $J$.
\endProof

Recall now that the {\it dimension} $d(\r)$ of $\r$ is defined as the
square root of the minimal index of $\r$.

\begproposition{A.4} Let $\s$ be a finite index endomorphism of the
factor $M$ and $\r$ an irreducible subsector of $\s$. If $w$ is an
isometry in $(\r,\s)$, then $$ V_\r={\sqrt{{d(\s)}\over
{d(\r)}}}w^*Jw^*JV_\s. $$
If $\s=\oplus_{i=1}^N n_i\r_i$ is an irreducible decomposition of
$\s$ and for each $i$  $\{w^{(i)}_k,\ k=1,\dots n_i\}$ is an
orthonormal basis of isometries in $(\r_i,\s)$, then
 $$
 V_\s=\sum_{i=1}^N\sum_{k=1}^{n_i}{\sqrt{{d(\r_i)}\over {d(\s)}}}
 w^{(i)}_kJw^{(i)}_kJV_{\r_i}.\eqno (A.4)
 $$
\endproposition

\begProof We prove the second assertion that implies the first one.
Set $W$ equal to the right hand  side in $(A.4)$. The ranges
of the $w^{(i)}_k$'s are pairwise orthogonal and the coefficients
verifies $\sum_{i=1}^Nn_i{d(\r_i)\over d(\s)}=1$, thus $W$ is an
isometry and a direct verification shows that it implements $\s$.
Moreover $W$ commutes with $J$, thus Lemma A.3 shows that $W$
implements a left inverse $\Phi$ of $\s$. But $W$ also preserves the
natural cone ${\cal P}^\natural(M,\Q)$, because this is true for
each of its terms, thus $W$ is the standard implementation of $\Phi$
by Proposition A.2. It remains to show that $\Phi$ is the minimal
left inverse. Now a left inverse is determined by the state obtained
by restricting it to $(\s,\s)$. The value of $\Phi$ on the minimal
projection $w^{(i)}_k w^{(i)*}_k$ is  ${d(\r_i)}\over {d(\s)}$,
hence it is the minimal left inverse. \endProof

\titlea{Appendix B. Invariant vectors for representations of
SL(2,$\Re$).}

We start by recalling the ``vanishing of the matrix coefficient
theorem'' for a  connected simple  Lie group $\bf G$ with finite
center, see [\rfr(Zimm1)].

\begtheorem{B.1 }Let $U$ be a unitary representation of $\bf G$
on a Hilbert space $\H$. If $U$ does not contain the identity, then
$(U(g)\x,\eta)\to 0$ as $g\to\infty$ for all $\x,\eta\in\H$.
\endtheorem

As a consequence, if $U$ is a unitary representation of $\bf G$ and
$\x\in\H$ then the subgroup $\{g\in{\bf G},\,U(g)\x=\x\}$ is either
compact or equals to $\bf G$.

In the following $\bf G$ always denotes the universal covering
group of $SL(2,\Re)$ and we state an explicit corollary in this case.
Let us consider the one-parameter subgroups of $\bf G$ of
the translations, dilations and rotations defined as the lifting
to $\bf G$ of the one-parameter subgroups of $SL(2,\Re)$
 $$
T(t)=\left( \matrix{1&t\cr0&1}\right),\quad
\Lambda(s)=\left(\matrix{e^{s\over2}&0\cr
0&e^{-{s\over2}}}\right),\quad
R(\th)=\left(\matrix{{\rm cos}{\th\over2} &{\rm sin}{\th\over2}\cr
-{\rm sin}{\th\over2} &{\rm cos}{\th\over2}}\right),\eqno(B.1)
 $$
 and we still denote them  by the same symbols $T,\Lambda,R$ (cf. the
definitions in Section~1).

 \begcorollary{B.2} Let $U$ be a unitary representation of $\bf G$
and $\Q$ a vector of the Hilbert space $\H$. The following are
equivalent:
 \item{$(i)$} $\Co\Q$ are the only $U$ invariant vectors.
 \item{$(ii)$} $\Co\Q$ are the only $U(T(\cdot))$ invariant
vectors.
 \item{$(iii)$} $\Co\Q$ are the only $U(\Lambda(\cdot))$ invariant
vectors.
\np
If moreover the generator of $U(R(\cdot))$ is positive then the
former are also equivalent to
 \item{$(iv)$} $\Co\Q$ are the only $U(R(\cdot))$ invariant
vectors.
\endcorollary

\begProof Although the cardinality of the center $\bf Z$ of $\bf G$
is infinite, we check that Theorem B.1 still applies. By decomposing
$U$ into a direct integral of irreducible representations, it is
sufficient to consider the case in which $U$ is irreducible. Since
$U$  is infinite-dimensional, the  tensor product  with its conjugate
representation $U\otimes\bar U$ does not contain the identity.
Now $U\otimes\bar U$ is trivial on the center $\bf Z$, hence
defines a representation of $PSL(2,\Re)$. If $\x\in\H$ then by
Theorem B.1
$$
|(U(g)\x,\x)|^2=(U(g)\otimes\bar
U(g)\x\otimes\bar\x,\x\otimes\bar\x)\to 0\quad{\rm as}\quad
g\to\infty. $$
Then the first set of equivalences is then clear. Furthermore
 $(i)$ is equivalent to $(iv)$ if the conformal Hamiltonian is
positive because the identity is the only irreducible unitary
representation of $\bf G$ with lowest weight $0$.
 \endProof

In this paper we need  a result in the spirit of Theorem B.1
concerning representations of the subgroup $\bP$ of the upper
triangular matrices in $SL(2,\Re)$, namely the group generated by
the translations and the dilations.

 \begproposition{B.3} Let $U$ be a unitary representation of $\bf
P$ on a Hilbert space $\H$. If $F\subset\H$ is a finite-dimensional
subspace which is globally $U(\Lambda(\cdot))$-invariant, then $F$
is left pointwise fixed by $U(T(\cdot))$.
 \endproposition

\begProof Setting $u(t):=U(T(t))$ and $v(s):=V(\Lambda(s))$ we have
two one-parameter unitary groups on $\H$ satisfying the commutation
relations
 $$
v(s)u(t)v(-s)=u(e^{s}t)\ , \qquad t,s\in\Re. \eqno(B.2)
 $$
Since $F$ is finite dimensional, we need to show that $u(t)\x=\x$ if
 $\x$ is a $v$-eigenvector, i.e. there exists a character
$\chi\in\widehat\Re$ such that
 $$
v(s)\x=\chi(s)\x\ ,\quad s\in \Re\ .\eqno(B.3)
 $$
 Indeed in this case by the formula $(B.2)$ implies
 $$
u(e^st)\x=v(s)u(t)v(-s)\x=\ov{\chi(s)}v(s)u(t)\x
 $$
hence
 $$
(u(e^st)\x,\x)=(u(t)\x,\x),\quad t,s\in\Re.
 $$
 As $s\to-\infty$ we thus have
 $$
(\x,\x)=(u(t)\x,\x)
 $$
that implies
$u(t)\x=\x$ by the limit case of the Schwartz inequality.
 \endProof
 Before concluding this appendix, we
recall a known fact needed in the text.

 \beglemma{B.4} Let $U$ be a unitary representation of $\bf G$.
The following are equivalent:
\item{$(i)$} The generator of $U(R(\cdot))$ is positive.
\item{$(ii)$} The generator of $U(T(\cdot))$ is positive.
\np
 In this case, if $U$ is non-trivial,  the spectrum of the
generator of  $U(T(\cdot))$ is $[0,\infty)$.\endlemma

\begProof For the equivalence $(i)\Leftrightarrow (ii)$  see e.g.
[\rfr(PrSe1)]. The last statement follows because the spectrum of
$U(T(\cdot))$ has to be dilation invariant because of the
commutation relations $(B.2)$. \endProof

\references

\end